\documentclass[12pt,preprint]{aastex}

\usepackage{color}
\usepackage{psfig}

\shorttitle{Critique of the X-wind Model}
\shortauthors{Desch et al.}

\begin{document}

\title{A Critical Examination of the X-Wind Model for Chondrule and 
Calcium-rich, Aluminum-rich Inclusion Formation and Radionuclide Production}

\author{S.~J.~Desch}

\and 

\author{M.~A.~Morris}

\affil{School of Earth and Space Exploration, Arizona State University,
        P.~O.~Box 871404, Tempe, AZ 85287-1404}
\and

\author{H.~C.~Connolly, Jr.}
\affil{Kingsborough Community College and the Graduate Center of the City University of New York,
                                2001 Oriental Boulevard, Brooklyn, NY 11235-2398}
\affil{American Museum of Natural History,
                                Central Park West at 79th Street, New York, NY 10024-5192}
\affil{Lunar and Planetary Laboratory, 1629 E. University Blvd.,
                                University of Arizona, Tucson, AZ 85721-0092}
\and

\author{Alan~P.~Boss}
\affil{Department of Terrestrial Magnetism, Carnegie Institution of Washington,
        5241 Broad Branch Road NW, Washington, DC 20015-1305}

\email{steve.desch@asu.edu}

\begin{abstract}
Meteoritic data, especially regarding chondrules and calcium-rich, aluminum-rich 
inclusions (CAIs), and isotopic evidence for short-lived radionuclides (SLRs) in 
the solar nebula, potentially can constrain how planetary systems form.  
Intepretation of these data demands an astrophysical model, and the ``X-wind" 
model of Shu et al.\ (1996) and collaborators has been advanced to explain the 
origin of chondrules, CAIs and SLRs.  It posits that chondrules and CAIs were 
thermally processed $<$ 0.1 AU from the protostar, then flung by a magnetocentrifugal
outflow to the 2-3 AU region to be incorporated into chondrites.  Here we critically 
examine key assumptions and predictions of the X-wind model.  We find a number of 
internal inconsistencies: theory and observation show no solid material exists
at 0.1 AU; particles at 0.1 AU cannot escape being accreted into the star; 
particles at 0.1 AU will collide at speeds high enough to destroy them; thermal 
sputtering will prevent growth of particles; and launching of particles in 
magnetocentrifugal outflows is not modeled, and may not be possible.  We also 
identify a number of incorrect predictions of the X-wind model: the oxygen fugacity 
where CAIs form is orders of magnitude too oxidizing; chondrule cooling rates 
are orders of magnitude lower than those experienced by barred olivine chondrules; 
chondrule-matrix complementarity is not predicted; and the SLRs are not produced 
in their observed proportions.  We conclude that the X-wind model is not 
relevant to chondrule and CAI formation and SLR production.  We discuss more 
plausible models for chondrule and CAI formation and SLR production.

\end{abstract} 

\section{Introduction} 

Chondrites, the most primitive known meteorites, are the witnesses to the birth of our
solar system. 
Their parent bodies, which are asteroids, formed some 4.57 billion years ago in the region
roughly 2-3 AU from the Sun 
and have suffered relatively little alteration since then (Wadhwa \& Russell 2000).
As such, they record conditions (chemistry, pressure, temperature) in the solar nebula.
Chondrites are the key to understanding our solar system's birth and, by extension, the 
processes in protoplanetary disks where planets are forming today. 

From a petrological standpoint, most chondrites are analogous to conglomerates, of 
igneous spheres.
Chondrites are remarkable for containing calcium-rich, aluminum-rich inclusions (CAIs),
the oldest solids formed in the solar system.
The formation of some CAIs has been dated very precisely: Pb-Pb dating of CAIs in the CV3 
chondrite NWA 2364 reveals an age $4568.67 \pm 0.17$ Myr (Bouvier \& Wadhwa 2009). 
The majority of CAIs (all but the Fluffy Type A, and related objects) experienced 
some degree of melting while floating freely in the solar nebula (Connolly et al.\ 2006).
Type B CAIs, in particular, were heated to high temperatures, followed by cooling over
periods of hours (at rates of $\approx \, 5 \, {\rm K} \, {\rm hr}^{-1}$; Stolper 1982; 
Stolper \& Paque 1986).
Also found in abundance within chondrites are sub-millimeter- to millimeter-sized,
(mostly ferromagnesian) igneous spheres, called chondrules.  
Chondrules formed at most $2-3 \, {\rm Myr}$ after CAIs (Amelin et al. 2002;
Kita et al. 2005; Russell et al. 2006; Wadhwa et al. 2007; Connelly et al. 2008), as 
melt droplets that were heated to high temperatures while they were independent, free-floating 
objects in the early solar nebula, after which they cooled over periods of hours.
Chondrules and CAIs together indicate that the components of chondrites were exposed to
widespread, energetic, transient heating events.

Unraveling the unusual process that melted chondrules and CAIs is fundamental to understanding 
the evolution of protoplanetary disks. 
That the mechanism was intermittent and transient follows directly from the inferred timescales 
for heating and cooling, which are hours or less.
The widespread nature of the mechanism is inferred from the fact that chondrules make up
as much as $\approx$ 80\% of the volume of ordinary chondrites (Grossman 1988).
The energies involved are staggering.
It is estimated that the current mass of chondrules in the asteroid belt is $\sim 10^{24} \, {\rm g}$
(Levy 1988).
The energy required to heat rock 1000 K and then melt it typically exceeds
$3 \times 10^{10} \, {\rm erg} \, {\rm g}^{-1}$, so at a minimum $3 \times 10^{34}$ ergs were
required to melt the existing chondrules.
For every gram of chondrules in the present-day asteroid belt, though, there were originally perhaps 
300 grams, subsequently lost as the asteroid belt was depleted by orbital resonances (Weidenschilling 1977a; 
Bottke et al.\ 2005; Weidenschilling et al. 2001).
As well, for every gram of rock in the solar nebula there was an associated 200 grams of gas
(Lodders 2003).
The energy to raise gas 1000 K in temperature exceeds $3 \times 10^{10} \, {\rm erg} \, {\rm g}^{-1}$
and thus far outweighs the energy needed to melt chondrules.
If chondrules were melted in the solar nebula and were thermally coupled to gas, the energy
required to heat the gas, along with all the chondrules inferred to have 
originally been there, exceeded $2 \times 10^{39} \, {\rm erg}$.
All in all, a remarkable fraction ($> 1\%$) of the gravitational potential energy of the disk mass
from 2 to 3 AU was involved in heating the gas during chondrule formation. 
In doing so, this mysterious mechanism did more than merely melt chondrules and CAIs: it left
clues to its nature in the manner in which chondrules and CAIs were melted, cooled and recrystallized.
The differences in igneous textures of chondrules and (type B) CAIs, combined with the elemental 
fractionations within crystals, provide constraints on their thermal histories (Connolly et al. 2006), 
and therefore provide constraints on the type of transient heating events that melted them.

Since Sorby (1877) first recognized the need to explain the igneous textures of chondrules, 
numerous mechanisms for melting of chondrules and CAIs have been proposed.
Some of the more favored mechanisms include interaction with the early active Sun, through jets 
(Liffman \& Brown 1995, 1996) or magnetic flares (Shu et al. 1996, 1997, 2001); melting 
by lightning (Pilipp et al.\ 1998; Desch \& Cuzzi 2000); melting by planetesimal impacts 
(Merrill 1920; Urey \& Craig 1953; Urey 1967; Sanders 1996; Lugmair \& Shukolyukov 2001); and also 
passage of solids through nebular shocks (
Wood 1963; Hood \& Horanyi 1991, 1993; Hewins 1997; Connolly \& Love 1998; Hood 1998; Jones et al. 2000; 
Iida et al.\ 2001; Desch \& Connolly 2002; Ciesla \& Hood 2002; Connolly \& Desch 2004; Desch et al.\ 2005; 
Connolly et al. 2006; Miura \& Nakamoto 2006; Morris \& Desch 2010). 
Of the proposed transient heating mechanisms, the two that have received the most attention
and which have been modeled in the most detail have been the nebular shock model and the so-called 
``X-wind model" of Shu et al. (1996, 1997, 2001).
The nebular shock model hypothesizes that chondrule precursors were overtaken by shocks passing through 
the gas of the solar nebula disk at about the present-day location of the chondrules, the asteroid
belt, 2-3 AU from the Sun. 
The source of these shocks may have been X-ray flares, gravitational instabilities, or bow shocks
driven by planetesimals on eccentric orbits (see Desch et al.\ 2005). 
Chondrules would be melted by the friction of the supersonic gas streaming past them, thermal exchange 
with the shocked, compressed gas, as well as by absorption of radiation from other heated chondrules. 
CAI precursors presumably formed in a hotter portion of the nebula but could have been melted by 
shocks as well. 
The X-wind model hypothesizes that solid material was transported to $< 0.1$ AU from the Sun, formed
chondrule and CAI precursors there, were melted, and then were transported back to 2-3 AU. 

Additional constraints on processes acting at the birth of the solar system arise from isotopic 
studies of meteorites, which reveal the presence of short-lived radionuclides (SLRs) in the solar nebula, 
radioactive isotopes with half-lives of millions or years or less. 
Although these isotopes have long since decayed, their one-time presence is inferred from excesses in
their decay products that correlate with the parent elements.
For example, the one-time presence of ${}^{26}{\rm Al}$, which decays to ${}^{26}{\rm Mg}$ with a
half-life of 0.71 Myr, is inferred by analyzing several minerals within a given inclusion, and finding
excesses in the ratio ${}^{26}{\rm Mg} / {}^{24}{\rm Mg}$ that correlate with the elemental ratio 
${}^{27}{\rm Al} / {}^{24}{\rm Mg}$.
The excesses are due to ${}^{26}{\rm Al}$ decay, so the proportionality between the ratios above yields
the value of ${}^{26}{\rm Al} / {}^{27}{\rm Al}$ when the inclusion crystallized (achieved isotopic
closure). 
In this way Lee et al.\ (1976) inferred an initial abundance
${}^{26}{\rm Al} / {}^{27}{\rm Al} \approx 5 \times 10^{-5}$ in CAIs from the carbonaceous chondrite
Allende.
Likewise several more SLRs have been inferred to exist, including such key isotopes as:
${}^{60}{\rm Fe}$ (Tachibana \& Huss 2003), with a half-life $t_{1/2} = 2.62$ Myr (Rugel et al.\ 2009); 
${}^{10}{\rm Be}$ (McKeegan et al.\ 2000), with $t_{1/2} = 1.5$ Myr;
and ${}^{36}{\rm Cl}$ (Lin et al.\ 2005), with $t_{1/2} = 0.36$ Myr.

The origins of these SLRs are debated, as reviewed by Wadhwa et al.\ (2007).
The consensus model, at least for the majority of SLRs, hypothesizes an origin in a nearby core-collapse
supernova, either just before or during the formation of the solar system.
Supernova material may have been injected into the Sun's molecular cloud core (Cameron \& Truran 1977;
Vanhala \& Boss 2002), or may have been injected into the Sun's protoplanetary disk (Chevalier 2000;
Ouellette et al.\ 2005, 2007). 
Indeed, the abundance of ${}^{60}{\rm Fe}$ is inconsistent with all models for its origin that do not
involve nearby, recent supernovae in the Sun's star-forming environment (Wadhwa et al.\ 2007). 
On the other hand, ${}^{10}{\rm Be}$ is not formed significantly in supernovae, and must have an
origin distinct from ${}^{60}{\rm Fe}$; this interpretation is supported by the observed decoupling 
of these two SLRs in meteorites (Marhas et al.\ 2002). 
Desch et al.\ (2004) point out that the abundance of Galactic cosmic rays (GCRs) that are themselves 
${}^{10}{\rm Be}$ nuclei is much higher than the ratio in the solar nebula, and that GCRs trapped
in the Sun's collapsing molecular cloud core will easily lead to the observed meteoritic abundance 
of ${}^{10}{\rm Be}$.
We discuss this model in somewhat more detail in \S 6.3. 
An alternative model for the origins of the SLRs is that they were created when energetic 
($> \, {\rm MeV} \, {\rm nucleon}^{-1}$) ions collided with nuclei of rock-forming elements brought 
$< 0.1$ AU from the Sun, in the context of the X-wind model (Gounelle et al.\ 2001). 
If this were true, a supernova source for the SLRs would not be demanded (except for ${}^{60}{\rm Fe}$).
Unraveling the origins of the SLRs has obvious, fundamental implications for where the Sun formed. 

The formation of chondrules and CAIs, and the origins of the SLRs, place important constraints on
the place of the Sun's origin, the presence of supernovae in its birth environment, and for processes in 
its protoplanetary disk.
These issues apply more broadly to protostars forming today, and bear on the likelihood of Earth-forming
planets. 
The X-wind model claims to explain chondrule and CAI formation, and the origins of the SLRs, in a 
unified model. 
The purpose of this paper is to critically examine the X-wind model. 
In \S 2, we first review the meteoritic constraints on the formation of chondrules and CAIs and on the 
origins of SLRs. 
We include petrologic constraints arising from the CAI {\it Inti} found in the {\it STARDUST} sample
return (Zolensky et al.\ 2006). 
The X-wind model itself is reviewed in \S 3. 
In \S 4 we discuss internal inconsistencies within the X-wind model, and in \S 5 we compare its 
predictions about chondrule and CAI formation and SLR production against the meteoritic constraints.
We discuss alternative hypotheses to the X-wind model in \S 6.
In \S 7 we draw conclusions about the viability of the X-wind model. 

\section{Meteoritic Constraints}

Isotopic and petrologic studies of chondrules and CAIs have yielded a wealth of
constraints about how these particles formed, and then were melted. 
Here we review the constraints that all models for the formation of chondrules and CAIs must satisfy. 
For further descriptions of these constraints, the reader is referred to reviews 
by Jones et al.\ (2000), Connolly \& Desch (2004), Desch et al.\ (2005), Connolly et al.\ (2006),
MacPherson (2003), and Ebel (2006). 
We also review the meteoritic evidence for SLRs and their possible origins. 
For further details, the reader is referred to Goswami \& Vanhala (2000), McKeegan \& Davis (2003), 
Gounelle (2006), and Wadhwa et al.\ (2007). 

\subsection{Chondrule Formation}

The most important constraints on chondrule formation come from experimental constraints
on their thermal histories.
Chondrules are the result of melting and recrystallization of precursor assemblages,
and constraints exist on the initial temperature of chondrule precursors, their peak 
temperatures and the time spent at these temperatures, as well as the cooling rates 
from the peak and during crystallization.
Here we highlight the main constraints only; the reader is referred to reviews on chondrule 
thermal histories by Desch \& Connolly (2002), Connolly \& Desch (2004), Desch et al.\ (2005),
and Hewins et al.\ (2005), and references therein. 
The initial temperatures of the chondrule precursors are generally held to be $< 650$ K,
the condensation temperature of S (at least in a solar-composition gas: Lodders 2003),
because chondrules contain primary S that was not lost during chondrule formation
(Rubin 1999; Jones et al. 2000; Tachibana \& Huss 2005; Zanda 2004).
Chondrules could not have spent more than a few hours at temperatures higher than 
$650 - 1200$ K, depending on pressure (Hewins et al.\ 1996; Connolly \& Love 1998; 
Rubin 1999; Jones et al.\ 2000; Lauretta et al.\ 2001; Tachibana \& Huss 2005). 
The majority of chondrules experienced peak temperatures in the range of 1770 - 2120 K  
for several seconds to minutes (Lofgren \& Lanier 1990; Radomsky \& Hewins 1990; Hewins 
\& Connolly 1996; Lofgren 1996; Hewins 1997; Connolly et al.\ 1998; Connolly \& Love 1998;  
Jones et al.\ 2000; Connolly \& Desch 2004; Hewins et al.\ 2005; Lauretta et al.\ 2006),
although the peak temperatures of barred olivine chondrules may have been as high as 2200 K 
(Connolly et al.\ 1998).  
Approximately 15\% of chondrules in ordinary chondrites contain relict grains (Jones 1996), 
whose survival depends on the time spent a chondrule spends at the peak temperature 
(Lofgren 1996; Connolly \& Desch 2004; Hewins et al.\ 2005).  
On this basis, chondrules spent only tens of seconds to several minutes at their peak 
temperatures (Connolly et al.\ 2006).
Likewise, retention of Na and S demands chondrules cooled from their peak temperatures at
rates $\sim 5000 \, {\rm K} \, {\rm hr}^{-1}$, or several hundred K in a few minutes
(Yu et al.\ 1995; Yu \& Hewins 1998).  
The textures of different chondrule textural types are reproduced experimentally only
by certain cooling rates through the crystallization temperature range (roughly
1400 - 1800 K for common chondrule compositions). 
In ordinary chondrites, 84\% of chondrules are pophyritic, with many euhedral crystals
(Gooding \& Keil 1981).  These are reproduced by cooling rates 
$\approx 5 - 1000 \, {\rm K} \, {\rm hr}^{-1}$ (Jones \& Lofgren 1993; Desch \& Connolly 2002).
Barred olivine textures, with many parallel laths of olivine, make up 4\% of ordinary
chondrite chondrules (Gooding \& Keil 1981), and require cooling rates 
$\approx 250 - 5000 \, {\rm K} \, {\rm hr}^{-1}$ (see Desch \& Connolly 2002 and references therein).
Finally, radial pyroxene textures, with a few crystals radiating from a single nucleation
site, account for 8\% of ordinary chondrite chondrules (Gooding \& Keil 1981).
These textures probably require destruction of relict grains and production of a supercooled
liquid (Connolly et al.\ 2006), and can be reproduced by cooling rates in the range
$5 - 3000 \, {\rm K} \, {\rm hr}^{-1}$ (Lofgren \& Russell 1986). 
Other chondrule textures exist, such as glassy chondrules that presumably cooled even faster
than these, but the salient point is that most chondrules were heated to temperatures
$> 1800 - 2000 \, {\rm K}$ for minutes only, cooling at $\sim 5000 \, {\rm K} \, {\rm hr}^{-1}$,
then cooled at slower rates $10^2 - 10^3 \, {\rm K} \, {\rm hr}^{-1}$ through their crystallization
temperatures $1400 - 1800 \, {\rm K}$. 

Besides these constraints on chondrule thermal histories during the chondrule-forming event,
other constraints restrict the timing of chondrule formation.
Chondrules contain relict grains, including unmelted fragments of large particles. 
The texture, chemistry, and oxygen isotopic composition of relict grains indicates that they are 
fragments of chondrules, formed in previous generations.
This signifies that the event that melted chondrules occurred more than once, and that 
individual chondrules may have experienced multiple heating events (Connolly et al.\ 2006; 
Ruzicka et al.\ 2008; Kita et al.\ 2008; Connolly et al.\ 2009).
From Al-Mg systematics, most extant chondrules are known to have melted approximately 2 Myr after 
CAIs formed (Russell et al.\ 1997; Galy et al.\ 2000; Tachibana et al.\ 2003; Bizzarro et al.\ 2004; 
Russell et al.\ 2006). 
These same data suggest timescales for chondrule formation of several Myr (Huss et al. 2001;
Tachibana et al. 2003; Wadhwa et al.\ 2007; Rudraswami et al.\ 2008; Hutcheon et al.\ 2009),
with 90\% formed between 1.5 and 2.8 Myr after CAIs (Villeneuve et al.\ 2009). 
U-Pb systematics confirm these timescales (Amelin et al. 2002; Kita et al.\ 2005; Russell et al.\ 
2006; Connelly et al.\ 2008) and, not incidentally, indicate that the Al-Mg system is a valid 
chronometer and that ${}^{26}{\rm Al}$ was homogeneously distributed in the solar nebula.

Finally, other constraints restrict the environment in which chondrules formed.
Chondrules almost certainly formed in the presence of dust that is to first order the 
matrix grains in which the chondrules are sited. 
Matrix in primitive carbonaceous chondrites contains forsterite grains that clearly
condensed from the gas and cooled at $\sim 10^{3} \, {\rm K} \, {\rm hr}^{-1}$ below 1300 K
(Scott \& Krot 2005). 
The similarity in cooling rate suggests that these matrix grains formed in the chondrule-forming
events. 
The cogenetic nature of matrix and chondrules is also strongly supported by the chondrule-matrix
chemical complementarity. 
Relative to a solar composition and to CI chondrites, all chondrules and matrix are depleted
in volatiles, even moderate volatiles, and metal-silicate fractionation leads to variable amounts 
of siderophile elements in chondrites; even the abundances of relatively refractory lithophiles 
(e.g., Ti, Ca, Al, Si, Mg and Fe) can vary within chondrules and matrix.
However, the bulk abundances of refractory lithophiles in many chondrites are closer to solar
abundances than the abundances of chondrules or matrix alone, strongly implying that the
chondrules and matrix grains {\it within a given chondrite} formed in the same vicinity within 
the solar nebula (Palme et al.\ 1993; Klerner \& Palme 2000; Scott \& Krott 2005; Ebel et al.\ 2008; 
Hezel \& Palme 2008).
Hezel \& Palme (2008) analyzed the Ca/Al ratios in the matrix and chondrules of Allende and Y-86751, two 
chondrites almost identical in bulk composition.
They found the Ca/Al ratio in the matrix of Allende to be sub-chondritic and the ratio in the matrix 
to be super-chondritic, with the exact opposite true in Y-86751.  
Ca and Al would be difficult to redistribute on the parent body, strongly implying that the chondrules
and matrix grains within these two chondrites formed from the same batch of material with near-solar 
composition; the two batches underwent slightly different degrees of fractionation of Ca and Al to form
one set of chondrules and matrix in Allende, and another set of chondrules and matrix in Y-86751.
The cogenetic nature of chondrules and matrix within a given chondrite means that the chondrite did not
form from very different reservoirs of material separated by time and place in the nebula, but in a 
particular time and place in the solar nebula, from solar-composition material, ostensibly near where 
chondrites originate today.

The density of chondrules in the chondrule forming region can be estimated as well. 
Cuzzi \& Alexander (2006) have investigated the lack of volatile loss from chondrules, which 
strongly implies high vapor pressures of volatiles in the chondrule forming region. 
So that evaporated volatiles remained in the vicinity of chondrules, the volume of 
gas per chondrule must not exceed $\sim 0.1 \, {\rm m}^{3}$ or, equivalently, the chondrule 
density was $> 10 \, {\rm m}^{-3}$.
So that volatiles not diffuse away from the chondrule-forming region, the chondrule-forming
region must have been $> 10^2 - 10^3 \, {\rm km}$ in extent. 
In addition, about 2.4\% of chondrules in ordinary chondrites are compound, stuck to another 
chondrule while semi-molten (Wasson et al.\ 1995). 
If chondrules had relative velocities $< 0.1 \, {\rm km} \, {\rm s}^{-1}$ (to avoid 
shattering upon impact) and were sufficiently plastic to stick for $\sim 10^{4} - 10^{5} \, {\rm s}$,
then the number density of chondrules ($\approx 300 \, \mu{\rm m}$ in diameter) must have
been $\approx \sim 0.1 - 1 \, {\rm m}^{-3}$ (Gooding \& Keil 1981), or 
$\sim 10 \, {\rm m}^{-3}$ if the relative velocities were $\sim 10 \, {\rm cm} \, {\rm s}^{-1}$, 
as implied by solar nebula turbulence models (Cuzzi \& Hogan 2003). 
For chondrules with masses $\approx 3 \, \times \, 10^{-4} \, {\rm g}$, these number densities
imply a mass density of chondrules $\approx 3 \times 10^{-9} \, {\rm g} \, {\rm cm}^{-3}$, 
larger than the nominal gas density, $\sim 10^{-9} \, {\rm g} \, {\rm cm}^{-3}$ (at 2-3 AU in a 
disk with 10 times the mass of the minimum mass solar nebula of Weidenschilling 1977a), and implies
that the solids-to-gas ratio was locally $\sim 300$ times greater than the canonical 1\%.
The enhancement of the solids-to-gas ratio is supported by the inference that the chondrule formation 
region was also relatively oxidizing.  
FeO-rich chondrules clearly formed in a gas much more oxidizing than one of a solar composition 
(Jones et al.\ 2000; Connolly \& Desch 2004; Fedkin et al.\ 2006). 
Possibly the elevated oxidation is due to chondrule vapor and/or evaporation of fine dust or water ice 
also concentrated in the chondrule-forming region (Fedkin et al.\ 2008; Connolly \& Huss 2010).
On the other hand, the solids-to-gas ratio may have been highly variable: FeO-poor chondrules apparently
formed in a more reducing environment, perhaps one as reducing as a solar-composition gas 
(Zanda et al.\ 1994; Jones et al.\ 2000; Connolly \& Desch 2004), although this interpretation is 
complicated by the possibility of reducing phases in the precursor assemblage such as C, so that 
chondrules may not so faithfully record the oxygen fugacity of the chondrule formation region 
(Connolly et al.\ 1994; Hewins 1997). 

One last, important constraint is the observed correlation between chondrule textures and compound 
chondrule frequency.
In ordinary chondrite chondrules overall, among the population of porphyritic, barred olivine and
radial pyroxene textures, 87\% are porphyritic, 4\% are barred, and 9\% are radial (Gooding \& Keil 1981).
Among compound chondrules in ordinary chondrites, which account for 2.4\% of all chondrules,
the proportions are 19\% porphyritic, 32\% barred, and 49\% radial (Wasson et al.\ 1995).
Barred olivines and radial pyroxenes are about an order of magnitude more common among compound 
chondrules than chondrules overall.  
Despite the rarity of compound chondrules, 24\% of barred olivines and 15\% of radial pyroxenes are
found in the compound chondrule population. 
Porphyritic textures are consistent with cooling rates $5 - 1000 \, {\rm K} \, {\rm hr}^{-1}$, although
chemical zoning profiles favor lower cooling rates (Jones \& Lofgren 1993; Desch \& Connolly 2002), 
while barred textures are reproduced only with cooling rates $250 - 3000 \, {\rm K} \, {\rm hr}^{-1}$.
The barred olivine textures that so strongly correlate with compound chondrules appear to require 
faster cooling rates, by about an order of magnitude (the cooling rates of radial pyroxenes are not
well determined, but appear to have been similarly fast).
These data strongly imply that chondrule cooling rates were faster where compound chondrules were 
more likely to form.  
If the solids-to-gas ratio varied in space, compound chondrules would have formed in regions of 
higher chondrule density. 
A positive correlation between chondrule cooling rate and chondrule density is then strongly implied. 

\subsection{CAI Formation} 

CAIs have long been recognized to be the assemblages of very refractory minerals such as hibonite, 
anorthite, spinel, perovskite and fassaite and high-temperature reaction products such as gehlenite 
and melilite that are the first to form from a cooling solar-composition gas (Larimer 1967; Grossman 1972; 
Ebel \& Grossman 2000).
These minerals are likely to have condensed out of a solar-composition gas as it cooled below 1800 K
(MacPherson 2003; Ebel 2006).
The site of this condensation is unknown: it may have occurred near the Sun, or in a transiently 
heated region farther away.
That the gas in the condensation region was of solar composition is supported not just by the
mineralogy of CAIs but by constraints on the oxygen fugacity of the CAI formation environment. 
The valence state of Ti (i.e., the Ti$^{4+}$/Ti$^{3+}$ ratio) in minerals such as fassaite 
and rh\"{o}nite in CAIs, which is sensitive to the $f{\rm O}_{2}$ during formation,
routinely show that CAIs formed in an environment with oxygen fugacity very near that of 
a solar composition gas, with $f{\rm O}_{2} \approx {\rm IW} - 6$, or 6 orders of magnitude less 
oxidizing than the Iron-Wustite buffer (Beckett et al.\ 1986; Krot et al.\ 2000; Simon et al.\ 2010; 
Paque et al.\ 2010).  
Recently, the mineral osbornite [(Ti,V)N] has been detected in two CAIs: a CAI within the CB chondrite
Isheyevo (Meibom et al.\ 2007), and the object known as {\it Inti} collected in the 
{\it STARDUST} sample return (Zolensky et al.\ 2006).
Significantly, osbornite can only condense in a gas that is very close in composition and oxidation
state to a solar-composition gas, with C/O ratios in the range 0.91 - 0.94 (Ebel 2006; Petaev et al.\ 2001). 
It is not possible to condense osbornite in an environment as oxidizing as that associated with 
chondrule formation, for example. 

Most CAIs were melted some time after their minerals condensed and the CAIs formed, but some CAIs
(the ``Fluffy Type A" CAIs) did not. 
For one class of melted CAIs (type B), peak temperatures $\approx$ 1700 K are inferred from the 
crystallization of melilite (Stolper 1982; Stolper \& Paque 1986; Beckett et al. 2006).  
Based on the inhomogeneous concentrations of V, Ti, and Cr within spinel grains, they are 
constrained to be at these peak temperatures for less than a few tens of hours (Connolly \& Burnett 2003).  
The cooling rates of Type B CAIs have been constrained to $0.5 - 50 \, {\rm K} \, {\rm hr}^{-1}$ 
(Paque \& Stolper 1983; MacPherson et al.\ 1984; Simon et al.\ 1996).  
Like chondrules, Type B CAIs show such petrographic and geochemical evidence for multiple heating events,
including variations in minor element concentrations in spinels and Na content in melilites (Davis \& 
MacPherson, 1996; Beckett et al., 2000, 2006; Connolly \& Burnett, 2000; Connolly et al.\ 2003). 
According to Beckett et al.\ (2000), after melting, some CAIs experienced alteration in the nebula before 
being re-melted. 
The time of such alteration is still unconstrained, but is clearly less than 1 Myr (Kita et al.\ 2005, 
2010: MacPherson et al.\ 2010). 
The overall timescale of CAI production has been constrained form the inferred initial abundance of 
${}^{26}{\rm Al}$ to be $\sim 10^{5}$ years (Young et al., 2005; Shahar \& Young, 2007; Kita et al., 2010; 
MacPherson et al., 2010), suggesting that the processing of refractory materials into igneous rocks was 
relatively rapid and stopped before chondrules were formed (Connolly et al., 2006).  
Thus, the processing of CAIs within the disk was cyclic over a relatively short time period of at most 
a few $\times 10^{5}$ years, but most likely $< 10^{5}$ years (Kita et al., 2010) 

Like chondrules, CAIs (at least, those of type B) experienced similar peak temperatures and cooling rates, 
and multiple melting events.
Unlike chondrules, CAIs equilibrated with a reducing gas with near-solar composition.
Their formation also occurred earlier in the nebula's evolution.
A reasonable interpretation is that CAIs formed earlier and were melted by a mechanism similar to that
that melted chondrules, but that CAIs were melted under different environmental conditions. 

\subsection{Short-Lived Radionuclides}

At this time, there are 9 SLRs with half-lives of $\sim 10^7$ yr or less that are inferred
from meteorites to have existed in the early solar system. 
The list of these SLRs, taken from the review by Wadhwa et al.\ (2007), is given in Table 1. 
\begin{deluxetable}{lcll}
\tablecolumns{4}
\small
\tablewidth{0pt}
\tablecaption{Short-lived radionuclides in the early solar system}
\tablehead{\colhead{Parent Isotope} & \colhead{$\;\;\;$T$_{1/2}$\tablenotemark{a}$\;\;\;$} 
					& \colhead{Daughter Isotope} & \colhead{Solar System Initial Abundance}}
\startdata
$^{41}$Ca&  0.1  &$^{41}$K   &$^{41}$Ca/$^{40}$Ca $\approx$ 1.5 x 10$^{-8}$ \\
$^{36}$Cl&  0.3  &$^{36}$Ar(98.1\%) &$^{36}$Cl/$^{35}$Cl $\approx$ 1.6 x 10$^{-4}$ ? \\
      &          &$^{36}$S(1.9\%)  &                                 \\
$^{26}$Al&  0.72 &$^{26}$Mg  &$^{26}$Al/$^{27}$Al $\approx$ 5.7 x 10$^{-5}$ \\
$^{60}$Fe&  1.5  &$^{60}$Ni  &$^{60}$Fe/$^{56}$Fe $\approx$ 3-10 x 10$^{-7}$ \\
$^{10}$Be&  1.5  &$^{10}$B   &$^{10}$Be/$^{9}$Be $\approx$ 10$^{-3}$ \\
$^{53}$Mn&  3.7  &$^{53}$Cr  &$^{53}$Mn/$^{55}$Mn $\approx$ 10$^{-5}$ \\
$^{107}$Pd& 6.5  &$^{107}$Ag &$^{107}$Pd/$^{108}$Pd $\approx$ 5-40 x 10$^{-5}$ \\
$^{182}$Hf& 8.9  &$^{182}$W  &$^{182}$Hf/$^{180}$Hf $\approx$ 10$^{-4}$ \\
$^{129}$I& 15.7  &$^{129}$Xe &$^{129}$I/$^{129}$Xe $\approx$ 10$^{-4}$ \\
\enddata
\tablenotetext{a}{Half-life in millions of years.}
\end{deluxetable}
The longest lived of these isotopes may have been continuously created over Galactic history
and inherited from the Sun's molecular cloud. 
Radionuclides are created by a variety of stellar nucleosynthetic processes, including
core-collapse supernovae, type Ia supernovae, novae, and outflows from Wolf-Rayet stars
and asymptotic-giant-branch (AGB) stars (Wadhwa et al.\ 2007). 
These are injected into the interstellar medium at a given rate and subsequently decay.
To the extent that the newly created isotopes are injected into the hot phase of the interstellar 
medium, incorporation of the SLRs into a forming solar system will only occur after the gas 
cools and condenses into molecular clouds.
This process, during which the gas remains isotopically ``isolated," takes considerable time, 
probably $\sim 10^8$ yr.
Recently Jacobsen (2005) and Huss \& Meyer (2009) have included such an isolation time in simple
Galactic chemical evolution models, and have used them to predict the abundances of SLRs inherited
from the interstellar medium.
Whether or not such intermediate-lived SLRs as ${}^{53}{\rm Mn}$, ${}^{107}{\rm Pd}$ and 
${}^{182}{\rm Hf}$ were inherited is debatable and dependent on input parameters.
A substantial fraction of ${}^{129}{\rm I}$ appears to be mostly inherited from the 
interstellar medium.
In fact, the solar nebula would have far too much of this SLR unless the isolation time
exceeds 100 Myr (Huss \& Meyer 2009).   
Inheritance of ${}^{53}{\rm Mn}$ at meteoritic abundances, however, is not possible with an
isolation time longer than $\sim 50$ Myr, so this isotope was probably not inherited. 
One robust finding of these studies is that even with a very short isolation time, inheritance
from the interstellar medium cannot yield the meteoritic abundances of ${}^{41}{\rm Ca}$, 
${}^{36}{\rm Cl}$, ${}^{26}{\rm Al}$ and ${}^{60}{\rm Fe}$.
These four SLRs, and probably ${}^{53}{\rm Mn}$ (and ${}^{10}{\rm Be}$ for that matter),
are diagnostic of a late addition to the solar nebula.

Since the X-wind models were published, strong evidence has arisen for the presence of 
live ${}^{36}{\rm Cl}$ ($t_{1/2} = 0.3 \, {\rm Myr}$) in the solar nebula, from Cl-S systematics 
of sodalite in carbonaceous chondrites, at levels ${}^{36}{\rm Cl} / {}^{35}{\rm Cl} \sim 4 \times 10^{-6}$ 
(Lin et al.\ 2005; Hsu et al.\ 2006), corroborating earlier hints from Cl-Ar systematics (Murty et al.\ 1997).
As sodalite is thought to be a late-stage product of aqueous alteration, the initial
${}^{36}{\rm Cl} / {}^{35}{\rm Cl}$ value would have been higher if it were injected by a supernova early 
in the nebula's evolution along with other SLRs.  
An initial value ${}^{36}{\rm Cl} / {}^{35}{\rm Cl} \sim 10^{-4}$ is usually inferred 
(Hsu et al.\ 2006; Wadhwa et al.\ 2007). 
More recent analyses of Cl-S systematics in wadalite in the Allende carbonaceous chondrite
indicate an even higher ratio,
${}^{36}{\rm Cl} / {}^{35}{\rm Cl} \approx 1.72 \pm 0.25 \times  10^{-5}$, implying even
higher initial abundances of ${}^{36}{\rm Cl}$ (Jacobsen et al.\ 2009).
These levels are higher than those thought possible for supernova injection, 
${}^{36}{\rm Cl} / {}^{35}{\rm Cl} \sim 10^{-6}$ (see discussion in Hsu et al.\ 2006), 
and have been interpreted as evidence for a late stage of irradiation within
the solar nebula, producing ${}^{36}{\rm Cl}$ by direct bombardment of target nuclei by 
energetic ions (Lin et al.\ 2005; Hsu et al.\ 2006; Jacobsen et al.\ 2009).
At this point it seems likely that this interpretation is correct, although the time and
place in the solar nebula where this irradiation took place are unknown.  
An irradiation origin of ${}^{36}{\rm Cl}$ does not necessarily imply an irradiation origin
within the X-wind environment. 

It is worth noting that Chaussidon et al.\ (2006) claimed evidence for the one-time 
presence of ${}^{7}{\rm Be}$, which decays to ${}^{7}{\rm Li}$ with a half-life of only 57 days, 
in a CAI from the carbonaceous chondrite Allende.
Li is notoriously mobile and subject to large isotope fractionations by chemical processes.
It is very difficult to distinguish radiogenic excesses of ${}^{7}{\rm Li}$ for these reasons.
Desch \& Ouellette (2006) identified several weaknesses of the analysis of Chaussidon et al.\
(2006).
They conclude that while Li indeed appears anomalous in this Allende CAI, perhaps representing
an admixture with spallogenic Li, the data are not conclusive whatsoever with any Li being
the decay product of ${}^{7}{\rm Be}$.

\section{Description of the X-wind Model} 

The X-wind model originally was developed by Shu and collaborators (Shu et al.\ 1994a,b, 1995;  
Najita \& Shu 1994; Ostriker \& Shu 1995), to explain the collimated outflows from protostars.
The X-wind model is first and foremost a model of gas dynamics in protostellar systems, and was 
extended only later to investigate the formation of chondrules and CAIs near the protostar, by 
Shu et al.\ (1996), Shu et al.\ (1997), and Shu et al.\ (2001), and to investigate nuclear processing 
of solids, by Lee et al.\ (1998) and Gounelle et al.\ (2001).
(See also reviews by Shu et al.\ 2000, Shang et al.\ 2000.)
We note that the model evolved somewhat through the late 1990s; we consider the models of 
Shu et al.\ (1996, 2001) for the dynamics and thermal processing of solids, and 
Gounelle et al.\ (2001) for the irradiation products, to represent the most recent and most 
detailed incarnations of the model.

\subsection{Dynamics}

We begin by summarizing the dynamics of gas and solids in the X-wind model.
At its heart, the X-wind is a magnetocentrifugal outflow, as in the classic work
of Blandford \& Payne (1982).
Magnetic field lines are anchored by flux freezing in the protoplanetary disk and
forced to co-rotate with it.
As they are whipped around by the disk, the inertia of matter tied to the field
lines causes the field lines far above and below the disk to bow outwards.
Ionized gas tied to the field lines acts like a bead on a wire: as the field line
(wire) is whipped around, the gas (bead) is flung outward.
This outflow carries significant angular momentum with it, and gas and entrained
solids accrete through the disk.

Gas and solids accrete until they reach the ``X point" at a distance $R_{\rm x}$ from
the protostar.
At the X point, the pressure of the stellar magnetic field prevents the inward flow
of disk gas and truncates the disk.
The value of $R_{\rm x}$, given by Equation 1 of Shu et al.\ (2001; see also Ghosh
\& Lamb 1979; Shu et al.\ 1994a) is easily reproduced under the assumption that the
magnetic pressure of the stellar magnetosphere balances the ram pressure of accreting
gas in the disk.
Outside $R_{\rm x}$, in the disk, gas is tied to open magnetic field lines that cross
the disk, and gas is driven outward by a magnetocentrifugal outflow.
Inside $R_{\rm x}$, magnetic field lines are tied to the protostar, and gas corotates
with the protostar.
Formally, the field lines and associated gas do not mix.
Shu et al.\ (1996, 2001) presume that ionization is low enough near the X point to allow
matter to diffuse across field lines and cross into the region interior to $R_{\rm x}$, but
this stage is not explicitly modeled.

As material crosses the X point, it is heated and expands along field lines.
Just farther than the X point, in the disk, where $T \approx 1500 \, {\rm K}$, the scale
height of the gas is $H \sim 2 \times 10^{10} \, {\rm cm}$ $\approx 0.03 \, R_{\rm x}$.
If the gas inside the X point is heated so that the scale height increases by a
factor of 30, the gas can flow directly onto the protostar, guided by the magnetic
field lines in a ``funnel flow".
Heating of the gas to $\sim 10^6 \, {\rm K}$ is sufficient, and can occur due
to heating by X-rays generated by reconnection events interior to the funnel flow.
The region interior to the funnel flow, denoted the ``reconnection ring", from
$r \approx 0.75 \, R_{\rm x}$ to $R_{\rm x}$, is modeled as having reversed poloidal components
across the midplane (Ostriker \& Shu 1995), leading to frequent magnetic reconnection
events akin to solar flares.
Shu et al.\ (2001) identify this region as a possible source of a component of protostellar X-rays
such as those observed by Skinner \& Walter (1998).
From such observations they infer an electron density
$n_{\rm e} \approx 3 \times 10^{8} \, {\rm cm}^{-3}$ and temperatures
$T \approx 8 \times 10^{6} \, {\rm K}$ in the reconnection ring, yielding sound speeds
$v_{\rm T} \sim 400 \, {\rm km} \, {\rm s}^{-1}$, gas densities
$\approx 5 \times 10^{-16} \, {\rm g} \, {\rm cm}^{-3}$, and pressures
$P \sim 10^{-7} \, {\rm atm}$.
Indeed, for the X wind model to work, this region needs to be the site of frequent
magnetic flares, so that solids in this region are irradiated by energetic ions and
undergo nuclear processing.

As gas accretes inward past the X point and joins the funnel flow, Shu et al.\ (2001) hypothesize
that a fraction $F \sim 0.01$ of the solid material leaves the flow and enters the
reconnection ring.
This can occur, they say, if solids spiral inward within the disk into the reconnection
ring, or if they fail to be lofted by the funnel flow.
Once in the reconnection ring, the solid particles orbit at Keplerian speeds through
a gas that is corotating with the protostar, and so experience a constant headwind.
This causes particles to lose angular momentum and spiral in towards the protostar
in a matter of years.
They are lost unless the magnetosphere of the protostar fluctuates, periodically waning
so that the disk can encroach on the reconnection ring, sweep up the particles and
launch them in a magnetocentrifugal outflow.
If they can be launched by the outflow, there is the possibility that the particles
can land in the disk, depending on their aerodynamic properties (Shu et al.\ 1996).

\subsection{Thermal Processing}

Shu et al.\ (2001) do not explicitly calculate the thermal histories of particles in the X wind.
They do not, for example, calculate temperature-dependent cooling rates $dT/dt$ vs. $T$.
They do, however, cite two possible mechanisms for thermally processing particles.
While in the reconnection ring or the disk, magnetic flares are presumed to heat chondrules
and especially CAIs; but CAIs and chondrules are {\it last} melted by sudden exposure to
sunlight as they are lofted away from the disk.

While in the reconnection ring, proto-CAIs are repeatedly exposed to magnetic flares
that heat particles, mostly by impacts by energetic ions and absorption of X-rays.
Depending on the flare energy luminosity and the area over which it is deposited,
CAIs can be mildly heated, destroyed completely in ``catastrophic flares", or
heated to the point where just their less refractory minerals evaporate.
While in the reconnection ring, it is assumed CAI material repeatedly evaporates
and recondenses.
An important component of the X-wind model as put forth by Shu et al.\ (2001) is that heating of
proto-CAI material will usually allow evaporation of ferromagnesian silicate
material, but leave unevaporated more refractory Ca,Al-rich silicate material.
Without this core/mantle segregation, irradiation by energetic ions (discussed below)
overproduces ${}^{41}{\rm Ca}$ with respect to ${}^{26}{\rm Al}$.
Flares are also presumed to heat chondrules in the transition region between
the reconnection ring and the disk.
Here the calculation of temperatures is very much intertwined with the structure
of the disk and the relative heating rates due to flares and sunlight.

The other mode of heating, and the one causing chondrules and CAIs to melt for
the last time before isotopic closure, arises when these particles are lofted by the
magnetocentrifugal outflows, above the disk in which they reside.
The presumed densities of proto-CAIs in the reconnection ring are such that they will
form an optically thick, if geometrically thin, disk.
Because this optically thick disk absorbs starlight obliquely, its effective temperature
due to heating by starlight, $T_{\rm disk}$, is lower than the particle blackbody
temperature $T_{\rm BB} = (L_{\star} / 16\pi r^2 \sigma)^{1/4}$ at that radius (where
$L_{\star}$ is the stellar luminosity and $\sigma$ the Stefan-Boltzmann constant).
Particles start within the disk at temperatures $\approx T_{\rm disk}$, but as they
are lofted their temperatures rise to $T_{\rm BB}$ as they are exposed to starlight.
Actually, they reach slightly higher $T$ because they are exposed to the radiation emitted
by the disk, as well; Shu et al.\ (1996, 2001) approximate this particle temperature, the highest
temperatures particles will reach, as $T_{\rm peak} \approx (T_{\rm disk}^{4} / 2 + T_{\rm BB}^{4})^{1/4}$.
For the parameters adopted by Shu et al.\ (1996, 2001) for the ``embedded" phase
(in which $\dot{M} \approx 2 \times 10^{-6} \, M_{\odot} \, {\rm yr}^{-1}$), 
we find $T_{\rm BB} \approx 1700 \, {\rm K}$, $T_{\rm disk} \approx 1160 \, {\rm K}$, and
$T_{\rm peak} \approx 1750 \, {\rm K}$ (approximately what Shu et al.\ 1996 find).
Thus, Shu et al.\ (1996) state that launching either a CAI or chondrule in an outflow can raise 
its temperature from $< 1200 \, {\rm K}$ to $1800 \, {\rm K}$ or more, within a span of ``a few hours." 
This timescale is set by the dynamics of the particle, which must travel roughly a scale
height in the vertical direction.
As the heated CAIs or chondrules are flung to great distances, the absorption of
starlight lessens, and they cool.
It is straightforward to demonstrate that the cooling rates in this scenario are necessarily
\begin{equation}
\frac{d T}{dt} \approx -\frac{1}{2} \frac{ v_{r} }{ r } \, T_{\rm BB}(r).
\end{equation}
For the trajectories depicted in Figure 2 of Shu et al.\ (1996), $v_{r} \approx 50 \, {\rm km} \, {\rm s}^{-1}$
at $r \approx 0.1 \, {\rm AU}$ where particles will cool through their crystallization temperatures.
This means that all particles---CAIs and chondrules---necessarily cool from their peak temperatures
at the same rate, about $10 \, {\rm K} \, {\rm hr}^{-1}$.

\subsection{Radionuclide Production}

A final, major component of the X-wind model is the production of SLRs in CAIs.
The reconnection ring is the site of frequent magnetic reconnection events.
If these act like solar flares, they could accelerate hydrogen and helium ions to energies in
excess of 1 MeV/nucleon.
Gounelle et al.\ (2001) hypothesize that flares akin to solar ``gradual" flares and ``impulsive" 
flares will take place
in the ring, and that ions are accelerated with the same efficiency, relative to the X-ray luminosity,
as in the solar atmosphere.
The flux today of energetic ($E > 10 \, {\rm MeV} \, {\rm nucleon}^{-1}$) ions at 1 AU today is
roughly $100 \, {\rm cm}^{-2} \, {\rm s}^{-1}$, yielding an energetic particle luminosity
$L_{\rm p} \sim 0.09 \, L_{\rm x}$ (Lee et al.\ 1998).
Because T Tauri stars have X-ray luminosities, presumably from flares, roughly five orders of
magnitude greater
(Feigelson \& Montmerle 1999; Feigelson et al.\ 2007; Getman et al.\ 2008 and references therein),
the fluence of such particles over, say, 20 years, if concentrated into the reconnection ring with
area $\sim 10^{24} \, {\rm cm}^{2}$, would reach $\sim 2 \times 10^{19} \, {\rm cm}^{-2}$.
Flares more akin to gradual flares would accelerate mostly protons and alpha particles and
lead to an energetic particle spectrum $\propto E^{-2}$, while
flares akin to impulsive flares would accelerate a comparable number of ${}^{3}{\rm He}$ ions,
and lead to an energetic particle spectrum $\propto E^{-4}$.
Proto-CAI material in the reconnection ring is constantly bombarded by these energetic ions,
which can initiate nuclear reactions in the rocky material, creating new isotopes.

Gounelle et al.\ (2001) simultaneously model the production of several SLRs within the context 
of the X-wind model, attempting to match their initial abundances as inferred from meteorites.
They model the production of 4 isotopes in particular:
${}^{10}{\rm Be}$ ($t_{1/2} = 1.5 \, {\rm Myr}$),
${}^{26}{\rm Al}$ ($t_{1/2} = 0.7 \, {\rm Myr}$),
${}^{41}{\rm Ca}$ ($t_{1/2} = 0.1 \, {\rm Myr}$),
and ${}^{53}{\rm Mn}$ ($t_{1/2} = 3.7 \, {\rm Myr}$).
They also model production of the very long-lived isotopes ${}^{138}{\rm La}$ ($t_{1/2} > 10^{12} \, {\rm yr}$)
and ${}^{50}{\rm V}$ ($t_{1/2} \sim 10^{11} \, {\rm yr}$), on the grounds that these are not produced
in abundance by stellar nucleosynthesis.
Of course, these isotopes are so long-lived that they are not diagnostic of irradiation in the solar
nebula; they could have been produced by spallation in molecular clouds over Galactic history, for
example.
We therefore focus on the discussion in Gounelle et al.\ (2001) of 
${}^{10}{\rm Be}$, ${}^{26}{\rm Al}$, ${}^{41}{\rm Ca}$
and ${}^{53}{\rm Mn}$.
These are produced overwhelmingly (but not exclusively) by nuclear reactions of H and He ions with
O, Mg and Al, Ca, and Fe nuclei, respectively.

Among the first findings of Gounelle et al.\ (2001) is that uniform irradiation of the average composition of proto-CAIs
will result in orders of magnitude more ${}^{41}{\rm Ca}$, relative to ${}^{26}{\rm Al}$, than is
observed in CAIs.
They found no way to reconcile the production rates of these two isotopes by irradiation, unless
two conditions were met: Ca (the primary target for ${}^{41}{\rm Ca}$) were sequestered in a core; and
the thickness of a Ca-free mantle surrounding the core were sufficiently thick to stop energetic ions.
Gounelle et al.\ (2001) assume that repeated evaporations of proto-CAIs preferentially leave behind a
residue of Ca,Al-rich refractory cores, onto which ferromagnesian silicates can condense.
Under these assumptions, Gounelle et al.\ (2001) found core sizes for which the meteoritic abundances of the 4 radionuclides
above were reproduced, to within factors of a few.

\section{Internal Inconsistencies of the X-wind Model} 

\subsection{Are Jets Launched by X Winds?}

Protostellar jets are virtually ubiquitous among protostars.
Moreover, jets are associated with strong magnetic fields and are apparently collimated by
magnetic hoop stresses (Ray et al.\ 2007). 
These observations strongly support models of protostellar outflows as
magnetocentrifugally launched.
They are also taken at times as support for the X-wind model in particular (Shu et al.\ 2000),
but it must be emphasized that jets could be taken as evidence for the X-wind
only if they can be shown to be launched from inside about 0.1 AU.
An ongoing debate in the astronomical community is whether protostellar jets
are launched from locations $\sim 0.1 \, {\rm AU}$ from the protostar,
as in the X-wind, or from $\sim 1 \, {\rm AU}$, as advocated by proponents of
``disk wind" models (Wardle \& K\"{o}nigl 1993; K\"{o}nigl \& Pudritz 2000; Pudritz et al.\ 2007).
To be blunt: just because one observes a protostellar jet and magnetocentrifugal
outflow from a disk does {\it not} mean that jets are launched from 0.1 AU,
let alone that solids in that disk are transported from a few AU, to 0.1 AU,
back out to a few AU.

In fact, the astronomical evidence at this time does not support the X-wind model,
and instead favors disk wind models.
Observations of radial velocities across jets reveal their angular momenta and
the launch point of the protostellar jets (Bacciotti et al.\ 2002; Anderson et al.\ 2003;
Coffey et al.\ 2004, 2007).
These observations are technically challenging and were only possible when the
{\it Hubble Space Telescope} / Space Telescope Imaging Spectrograph ({\it HST}-STIS)
was operational.
Not all observations were successful; in some cases jet rotation was not observed.
In other cases rotation was observed, but in the opposite sense of the disk's presumed
rotation, complicating the interpretation (Cabrit et al.\ 2006; Pety et al.\ 2006;
Coffey et al.\ 2007; Lee et al.\ 2006, 2007).
Prograde jet rotation was observed in some protostellar systems, though; in those
systems jets appear to be launched from much farther in the disk than the X point.
Coffey et al.\ (2004) observed jet rotation in RW Aur and LkH$\alpha$321, and
more detailed observations were carried out by Coffey et al.\ (2007).
In DG Tau, a high-velocity component appears launched from about 0.2 - 0.5 AU and a
low-velocity component from as far as 1.9 AU; in TH 28, the jet seems launched from
about 1.0 - 3.9 AU; and in CW Tau, from 0.5-0.6 AU (Coffey et al.\ 2007).
These authors admit they have not resolved the innermost jet and cannot exclude
a contribution from an X-wind; but Woitas et al.\ (2005) estimate that the jets
carry at least 60-70\% of the angular momentum to be extracted from the disk.
Clearly the disk winds dominate in these systems.
As yet, there is no direct evidence from observations of jet rotation that outflows 
are launched by an X-wind rather than disk winds.

\subsection{Solids at the X point?} 

Besides the question of whether outflows are launched from inside 0.1 AU at all,
a second major obstacle for the X-wind model is that neither the model itself nor
astronomical observations support the existence of solids at the X point.
The theoretical grounds for a lack of solids at the X point are simple.
In calculating the temperature of disk material, Shu et al.\ (1996, 2001) neglected 
the heating of the disk due to its own accretion, focusing only on the passive heating 
of the disk by starlight.
Specifically, they set
\begin{equation}
\sigma T^{4}_{\rm disk} = \frac{L_{\star}}{4\pi^2 R_{\star}^{2}} \,
\left[ \arcsin \left( \frac{ R_{\star} }{ r } \right)
              -\left( \frac{ R_{\star} }{ r } \right) \,
 \left( 1 - \frac{ R_{\star}^{2} }{ r^{2} } \right)^{1/2} \right],
\label{eq:tempstar}
\end{equation}
which for parameters they consider typical of the embedded phase
($L_{\star} = 4.4 \, L_{\odot}$, $r = R_{\rm x} = 4 R_{\star} = 12 R_{\odot}$)
yields $T_{\rm disk} \approx 1160 \, {\rm K}$.
For parameters they consider typical of the revealed phase 
($L_{\star} = 2.5 \, L_{\odot}$, $r = R_{\rm x} = 5.3 R_{\star} = 16 R_{\odot}$),
$T_{\rm disk} \approx 820 \, {\rm K}$.
But an additional term must be added to the right side of Equation~\ref{eq:tempstar}
to account for energy released by disk accretion.
Setting
\begin{equation}
\sigma T_{\rm acc}^4 = \frac{3}{8\pi} \dot{M} \Omega^2,
\end{equation}
we can better estimate the effective temperature of the disk 
(approximately the temperature at optical depths $\approx 1$ into the disk's surface) as
\begin{equation}
\sigma T_{\rm eff}^{4} = \sigma T_{\rm disk}^{4} + \sigma T_{\rm acc}^{4},
\end{equation}
(Hubeny 1990).
Using $\Omega \approx 8 \times 10^{-6} \, {\rm s}^{-1}$ at the X point
and assuming a mass accretion rate of $2 \times 10^{-6} \, M_{\odot} \, {\rm yr}^{-1}$
for the embedded phase, one derives $T_{\rm acc} = 2030 \, {\rm K}$ and 
a temperature $T_{\rm eff} \approx 2090 \, {\rm K}$,
sufficient to evaporate all solids.
Even if one uses the lower mass accretion rate 
$\dot{M} \approx 1 \times 10^{-7} \, M_{\odot} \, {\rm yr}^{-1}$,
appropriate for the revealed stage, $T_{\rm acc} = 960 \, {\rm K}$ and 
$T_{\rm eff} \approx 1070 \, {\rm K}$.
The effective temperature is approximately the temperature at optical depths
$\approx 1$ into the disk's surface.

These high temperatures are exacerbated by the fact that $T_{\rm eff}$ is a lower 
limit to the temperatures experienced by particles.
The effective temperature is approximately the temperature of the disk at 1 optical
depth into the disk.
Because accretional heating must be transported out of the disk by a radiative flux,
temperatures inside the disk, at optical depths $\gg 1$ (using the Rosseland mean opacity)
will exceed $T_{\rm eff}$, by a factor $\approx (3\tau/8)^{1/4}$ (Hubeny 1990).
For even moderate optical depths (e.g., $\tau = 10$), temperatures will rise above 1500 K, 
even for the lower mass accretion rates of the revealed stage.
Considering optical depths $\ll 1$, temperatures will also exceed $T_{\rm eff}$, because
the particles will be exposed to starlight directly.
An isolated particle at a distance $r$ from the protostar will achieve a blackbody
temperature
\begin{equation}
T_{\rm BB} = \left( \frac{ L_{\star} }{16\pi \sigma r^2} \right)^{1/4}.
\end{equation}
For particles at the X point, $T_{\rm BB} \approx 1700 \, {\rm K}$ during
the embedded stage, and $\approx 1280 \, {\rm K}$ during the revealed stage.
In addition to the direct starlight, particles in the uppermost layers of the disk
will also absorb radiation from the disk as well, achieving temperatures well 
approximated by
\begin{equation}
T^{4} \approx \frac{1}{2} T_{\rm eff}^{4} + T_{\rm BB}^{4}
\end{equation}
(Shu et al.\ 1996).
Even for the revealed stage, this temperature is 1360 K.
Finally, if the dust particles in the uppermost layers are submicron in size, they
will absorb optical radiation but will be unable to radiate in the infrared effectively, 
and they will achieve even higher temperatures still. 
Chiang \& Goldreich (1997) have explained the excess near-infrared emission in
 spectral energy distributions (SEDs) of protostellar disks by accounting for this
``superheated" dust layer.
Even during the revealed stage, then, particles in the uppermost layers of the disk
at the X point will achieve temperatures in excess of 1360 K.
The significance of the dust temperatures $> 1360 \, {\rm K}$ is that silicates are 
not stable against evaporation such high temperatures (at least in the disk 
environment discussed here, mixed in a solar ratio with ${\rm H}_{2}$ gas).
Above 1400 K, for example, dust grains will evaporate in only hours (Morris \& Desch 2010). 
Thus, temperatures are simply too high to have a dusty disk approach all the way to 
the X point, even during the ``revealed" stage, when mass accretion rates are 
$\leq 10^{-7} \, M_{\odot} \, {\rm yr}^{-1}$.
A calculation of the innermost radius where dust can stably reside is complicated by 
the ``wall-like" structure of the disk there, and the poorly known thermodynamic
properties of dust materials, but has been considered by Kama et al.\ (2009), who show 
that typically the inner edge where solids can exist is typically several 
$\times \, 0.1 \, {\rm AU}$ from a protostar. 

Astronomical observations confirm the absence of solids at the X point.
Eisner et al.\ (2005) have determined the inner edges of dust emission in the
protoplanetary disks surrounding 4 Sun-like protostars, through a combination
of NIR interferometry and SED fitting.
Through measurements of other stellar properties, they also determined the
locations of the corotation radius and the predicted locations of the X point.
They find that typically the corotation radius and magnetospheric truncation
radius are both $< 0.1 \, {\rm AU}$ and agree within the uncertainties, but
that the inner edge of the dust disk also typically lies beyond either of these
radii, at about $0.1 - 0.3 \, {\rm AU}$.
This is true even for V2508 Oph, the protostar with the least discrepancy (among
the 4 sampled) between the X point and the inner edge of the dust disk.
It is also a protostar with parameters that closely match those adopted by
Shu et al.\ (1996, 2001) for a protostellar system in the revealed stage:
$M_{\star} = 0.9 \, M_{\odot}$, $\dot{M} = 2.3 \times 10^{-7} \, M_{\odot} \, {\rm yr}^{-1}$,
and an age $\approx 0.6 \, {\rm Myr}$.
Eisner et al.\ (2005) attribute the existence of an inner edge to the dust disk
to sublimation of dust at that radius, consistent with their observation that
the maximum temperature associated with dust emission is in the range 1000 - 2000 K
($\approx$ 1500 K for V2508 Oph, albeit with considerable uncertainty).
Eisner et al.\ (2005) also note that in systems with higher mass accretion rates,
the X point (by construction) is pushed inward, and they observed the inner edge
of the dust disk to move outward.
This finding is also consistent with dust sublimation being the cause of the inner
edge of the disk.
Based on the theoretical arguments above, and the observations of Eisner et al.\
(2005), solid particles are not expected to exist at the X point in disks with
mass accretion rates $> 10^{-7} \, M_{\odot} \, {\rm yr}^{-1}$.
Altogether, by neglecting accretional heating, Shu et al.\ (1996, 2001) appear to have
underestimated the temperatures of solids, and predicted them to exist where
they should not be and, indeed, are not observed to be.

\subsection{Decoupling from the Funnel Flow?}

In order for the X-wind model to be a valid description of CAI or chondrule formation, these 
objects must adhere to a specific dynamical history.
Specifically, Shu et al.\ (2001) assumed that a fraction $F \sim 0.01$ of all solid material
decouples from the funnel flow and enters the reconnection ring.
It is presumed to do so because it is bound in solid particles that experience a gravitational
force greater than the drag force exerted on them by the funnel flow. 
We argue above that all material should evaporate at the X point, but assuming solids to exist, 
their dynamical histories will depend critically on their sizes. 
Clearly protoplanetary disks contain sub-micron and micron-sized grains,
as evidenced by silicate emission features at $10 \, \mu{\rm m}$ (e.g., Sargent et al.\ 2009).
Shu et al.\ (1996, 2001) specifically identify these micron-sized solid particles
with matrix grains in chondrites.
Importantly, within the context of the X-wind model, there are no other particles in chondrites 
that can be identified as pre-existing in the protoplanetary disk, because chondrules and CAIs
form in the X-wind environment, and not in the disk. 
Chondrites also contain large aggregations of smaller particles that are unmelted, only lightly 
sintered and lithified, termed agglomeratic chondrules; but these are rare, making up only 2\% 
of the volume of ordinary chondrites (Dodd \& van Schmus 1971; Weisberg \& Prinz 1996). 
We discuss these below, but for now assert that if chondrules do not form in the disk, then
for practical purposes the only solid material entering the funnel flow would be micron-sized 
grains.

Because solid particles in the disk are so small, they are almost certain to couple
strongly to the gas as it enters the funnel flow.
According to Weidenschilling (1977b), small particles with aerodynamic stopping times
much less than the dynamical time will basically move with the gas, but with a small
relative velocity $(\Delta g) t_{\rm stop}$, where
\begin{equation}
t_{\rm stop} = \frac{ \rho_{\rm s} a }{ \rho_{\rm g} v_{\rm T} }
\end{equation}
is the aerodynamic stopping time [in the Epstein drag limit where particles
are smaller than the mean free path of gas molecules, appropriate for micron-sized
particles in gas with density $< 10^{-4} \, {\rm g} \, {\rm cm}^{-3}$,
or chondrules in gas with density $< 10^{-7} \, {\rm g} \, {\rm cm}^{-3}$],
where $\rho_{\rm s}$ and $a$ are the particle density and radius,
$\rho_{\rm g}$ and $v_{\rm T}$ are the gas density and thermal velocity,
and $\Delta g$ is the difference between the accelerations felt by the gas and solids.

In the context of the disk proper, $\Delta g$ is the extra acceleration the gas
feels because of pressure support,
\begin{equation}
\Delta g = \frac{1}{\rho_{\rm g}} \, \frac{\partial P_{\rm g}}{\partial r},
\end{equation}
where $P_{\rm g}$ is the gas density.
Assuming $T \approx 1500 \, {\rm K}$ just outside the X point,
$\Delta g \sim v_{\rm T}^{2} / r \sim 0.1 \, {\rm cm} \, {\rm s}^{-2}$
(neglecting terms of order unity).
The disk scale height is $H \sim 2 \times 10^{10} \, {\rm cm}$, and assuming
a minimum mass solar nebula (Weidenschilling et al.\ 1977a), we estimate a disk
density $\Sigma \sim 10^5 \, {\rm g} \, {\rm cm}^{-2}$ at the X point, yielding
a gas density $\rho_{\rm g} \sim 10^{-6} \, {\rm g} \, {\rm cm}^{-3}$.
For a particle with radius $a = 1 \, \mu{\rm m}$ and internal density
$\rho_{\rm s} = 3 \, {\rm g} \, {\rm cm}^{-3}$, the aerodynamic stopping time
is $t_{\rm stop} \sim 10^{-3} \, {\rm s}$.
The relative velocity between gas and dust, within the disk, is therefore
$\sim 10^{-4} \, {\rm cm} \, {\rm s}^{-1}$.
This relative velocity is negligible, and gas and dust can be considered
perfectly coupled.

In the context of the transition between the disk and the funnel flow,
$\Delta g$ is given by the acceleration the gas experiences.
Shu et al.\ (1996, 2001) do not explicitly model this stage, but we can estimate the
acceleration as follows.
The gas starts essentially from rest at the X point, but by the time it
participates in the funnel flow it could be moving as much as the thermal
velocity in the reconnection ring, $V \sim 400 \, {\rm km} \, {\rm s}^{-1}$.
The distance over which this occurs is perhaps
$d \sim 0.1 \, R_{\rm x} \sim 10^{11} \, {\rm cm}$.
Thus $\Delta g \sim V^{2} / d \sim 10^{4} \, {\rm cm} \, {\rm s}^{-2}$
(about 10 g's).
As for the stopping time, we derive a lower limit to the gas density in
the funnel flow by assuming that it carries a total mass flux $\dot{M}_{\star}$ onto
the star.
The funnel flow arises from an area $A$, and is composed of gas moving at a velocity
$V$, with density $\rho_{\rm g} = \dot{M}_{\star} / (A V)$.
The lower limit to the density is found by setting $A$ and $V$ as large as they
can be, and using the smallest value of $\dot{M}_{\star}$.
The absolute largest $A$ can be is $4\pi R_{\rm x}^{2} \sim 8 \times 10^{24} \, {\rm cm}^{2}$,
but the size of the reconnection ring, $\sim 1 \times 10^{24} \, {\rm cm}^{2}$, is probably
still an overestimate to the true value of $A$.
We take the thermal velocity of the gas (after heating to $10^{7} \, {\rm K}$),
$v_{\rm T} \sim 400 \, {\rm km} \, {\rm s}^{-1}$, to represent the maximum velocity of
the gas.
Thus $\rho_{\rm g} > 10^{-13} \, {\rm g} \, {\rm cm}^{-3}$ in the funnel flow, and
$t_{\rm stop} < 10^{2} \, {\rm s}$.
Micron-sized particles (or their aerodynamic equivalents) therefore reach relative
velocities with respect to the gas no more than
$\sim (\Delta g) t_{\rm stop} < 10 \, {\rm km} \, {\rm s}^{-1}$.
This velocity sounds significant [indeed, it would probably lead to evaporation of the dust
grains by frictional drag cf. Harker \& Desch (2002)] until it is remembered that it is only 
2\% of the total velocity: both gas and solid particles will move on nearly identical funnel-flow 
trajectories.
Over the roughly 1 hour ($= d / V$) the gas takes to accelerate from the disk to the
funnel flow, particles will be displaced only about $2 \times 10^9 \, {\rm cm}$
$= 0.002 \, R_{\rm x}$, a negligible amount.
Put another way, if gas is funneled onto one spot on the protostar, taking $\sim 10 \, {\rm hr}$
to reach it, dust grains will arrive 10 minutes later, at a spot about 1\% of the protostar's
radius away.

Shu et al.\ (2001) argue that solid particles can ``fall out" of the funnel flow if the gravitational force
on them exceeds the drag force lifting them.
This requires
\begin{equation}
\frac{4\pi}{3}\rho_{\rm s} a^{3} \, \Omega^2 \, z  >
\pi a^{2} \, \rho_{\rm g}  (C_{\rm D} / 2) \, V_{\rm g}^{2},
\end{equation}
where $z$ is the height above the midplane.
Taking $z \sim H \sim 2 \times 10^{10} \, {\rm cm}$ (the scale height of the disk),
and $C_{\rm D} = (2/3) (\pi k T_{\rm p} / \bar{m} )^{1/2} / V_{\rm g}$ (Gombosi et al.\ 1986),
the condition to fall out of the flow becomes a lower limit to the particle size:
\begin{equation}
a > a_{\rm crit} = \frac{1}{4 \rho_{\rm s} \Omega^{2} z} \,
\left( \frac{ \pi k T_{\rm p} }{ \bar{m} } \right)^{1/2} \,
\frac{ \dot{M}_{\star} }{ A },
\end{equation}
where the same relationship between mass accretion rate and gas density in the funnel flow
as above was used.
Taking $T_{\rm p} = 1500 \, {\rm K}$, $\bar{m} = 0.6 \, m_{\rm H}$, a mass accretion
rate $\sim 10^{-7} \, M_{\odot} \, {\rm yr}^{-1}$ and and area $\sim 10^{24} \, {\rm cm}^{2}$,
the critical particle diameter to fall out of the funnel flow is $\sim 4 \, {\rm mm}$, and is much
larger for higher mass accretion rates.

The conclusion to be reached from all this is that solid material accreting inward, from
the disk, through the X point, will remain coupled to the gas as it participates in a funnel
flow, unless the solid material in the funnel flow is aerodynamically equivalent to 
{\it compact} spheres, several millimeters in diameter.
Such particles cannot be chondrules and CAIs, since these are presumed not to form in the disk
in the X-wind model, and matrix grains are clearly too small to dynamically decouple from the
gas. 
Agglomeratic chondrules are larger than matrix grains, with diameters $0.3 - 1 \, {\rm mm}$ typically,
but that is still too small to decouple from the funnel flow.
This is true even if they were compact objects in the nebula gas, but models of coagulation predict 
that such aggregates would be fractal in shape (Dominik \& Tielens 1997).
It is quite possible these objects compacted only during accretion onto the parent body; if so, 
they would have behaved aerodynamically like the smallest particles of which they are composed, 
i.e., like micron-sized grains (Dominik \& Tielens 1997), making it even less likely that they could
have decoupled from the funnel flow.
Finally, the fact that agglomeratic chondrules make up only 2\% of the volume of ordinary
chondrites (Weisberg \& Prinz 1996), while chondrules make up 85\% (Gooding \& Keil 1981) is 
difficult to reconcile with the idea that chondrules and CAIs formed, with low efficiency, from 
such agglomerations.
Thus, there is no significant (i.e., at the $\sim 1\%$ level) component of solid material in the 
disk that can be expected to decouple from the funnel flow.
The assumption that a fraction $F \sim 0.01$ of all solid material would leave the funnel flow 
and enter the reconnection ring, an assumption Shu et al.\ (2001) themselves term ``{\it ad hoc},"
appears invalid.
Even if solid material existed at the X point, the fraction that would fall out of the funnel 
flow would be $\ll 0.01$.

\subsection{Survival and Growth in the Reconnection Ring?}

The arguments above suggest that solids would not decouple from the funnel flow.
Assuming anyway that solid material can enter the reconnection ring, we examine the dynamics
of particles there, and also their growth and survival.
Growth of solid material in the reconnection ring is much dependent on the dynamics of particles,
because the relative velocities $w$ between particles will determine the sticking coefficient $S$, the
probability that the two particles will stick rather than bounce off or even destroy each other.
Shu et al.\ (2001) note (after their Equation 31) that $w$ is implicitly assumed to be small enough that
``molten rocks stick rather than splatter on colliding".
The upper limit on $w$ obviously will depend on particle composition and whether it is
molten or solid, but a typical upper limit adopted in the literature on compound chondrules,
which are molten as they collide, is $\sim 0.1 \, {\rm km} \, {\rm s}^{-1}$
(e.g., Gooding \& Keil 1981; Ciesla \& Hood 2004).
Dominik \& Tielens (1997) calculate that solid particles will on average shatter if they
collide at velocities $> 0.01 \, {\rm km} \, {\rm s}^{-1}$.
In any plausible scenario, however, falling out of the funnel flow would impart vertical velocities 
to particles comparable to the Keplerian velocities, $\sim 10^2 \, {\rm km} \, {\rm s}^{-1}$,
essentially putting particles on orbits with different inclinations.
Necessarily, the relative velocities between particles will also be comparable to
these Keplerian velocities.
The gas drag forces acting on the particles in the reconnection ring are completely inadequate
to slow the incoming particles before they collide with and destroy particles already in the
reconnection ring (the surface density of gas, $\sim 10^{-5} \, {\rm g} \, {\rm cm}^{-2}$, will
not stop even micron-sized particles in less than dozens of disk crossings, while the optical 
depth of particles in the reconnection ring is large enough to ensure an impact with every crossing). 
Thus the actual relative velocities of colliding particles in the reconnection ring would exceed the 
shattering limit, by orders of magnitude.

Put another way, so that particles in the reconnection ring do not collide and shatter, they must 
exist in a very thin disk with low dispersion of relative velocities, $w_{z}$.
Defining, as Shu et al.\ (2001) do, $w_{z} \sim \alpha w$, where $\alpha \sim 0.3$, then the scale
height of the disk of proto-CAIs would have to be $H_{r} \sim w_{z} / \Omega \sim 3 \times 10^{8} \, {\rm cm}$
$\sim 10^{-2}$ times the scale height of the disk proper, in order for most particles
not shatter each other on impact.
As particles would overwhelmingly exit the funnel flow at much greater heights above the disk, it is
inevitable that they would not collect in the reconnection ring, but rather shatter upon impact there.

We calculate the effect of all of these particles falling out of the funnel flow as follows.
Assuming the mass flux in the funnel flow is $\dot{M} \sim 10^{-7} \, M_{\odot} \, {\rm yr}^{-1}$,
and a fraction $\sim 10^{-2}$ of that is in the form of solids, of which a portion $\sim 10^{-2}$
decouples from the funnel flow, then the flux of particles into the reconnection ring is 
$\sim 10^{-11} \, M_{\odot} \, {\rm yr}^{-1}$, or 
$\sim 2 \times 10^{22} \, {\rm g} \, {\rm yr}^{-1}$.
Spreading out this flux of particles over the area of the reconnection ring
$\sim 10^{24} \, {\rm cm}^{2}$, we estimate a solid particle flux 
$\sim 2 \times 10^{-2} \, {\rm g} \, {\rm cm}^{-2} \, {\rm yr}^{-1}$. 
A growing CAI has a radius $> 100 \, \mu{\rm m}$ and a cross section $\sim 3 \times 10^{-4} \, {\rm cm}^{2}$,
and so intercepts a mass $> 6 \times 10^{-6} \, {\rm g} \, {\rm yr}^{-1}$ from solid particles falling
out of the funnel flow, or $\approx 2 \times 10^{-4} \, {\rm g}$ over 30 years.
This mass exceeds by a large factor the mass of the growing CAI itself, so it is easy to see that
a growing CAI will collide with its own mass over its residence time in the disk, at speeds far exceeding 
tens of km/s.
This alone will prevent particles from growing in this environment.

Supposing anyway that the relative velocities are slow enough so that particles don't shatter, it still
is not clear that the sticking coefficient will be sufficient to allow growth.
Shu et al.\ (2001) suggest that $S$ might be low unless particles are molten, immediately following heating
by a flare.
Since flares have a limited extent and duty cycle, Shu et al.\ (2001) adopt an effective sticking coefficient
$S \sim 8 \times 10^{-4} \, (2\pi)^{1/2} \, \alpha$, or $S < 10^{-3}$, as typical.
To assume a higher value for $S$, particles would have to somehow stick even while completely solid. 
Shu et al.\ (2001) calculate the mass flux onto a particle as 
\begin{equation}
4\pi a^{2} \rho_{\rm s} \frac{d a}{dt}
\approx +\frac{3}{4 (2\pi)^{1/2}} \, \left( \Sigma_{\rm r} \Omega \right) \, \frac{S}{\alpha},
\end{equation}
where $\Sigma_{\rm r}$ is the assumed surface density of rock in the reconnection ring.
(NB: This appears to overestimate the growth rate by a factor $3(\pi/8)^{1/2} \sim 2$.) 
The important points about this formula are that the time rate of change of particle radius
is independent of radius, and that the growth rate is proportional to the surface density of 
rocky material, which only reaches a maximum value 
$\sim \Sigma_{\rm r} \sim 1.6 \, {\rm g} \, {\rm cm}^{-2}$ about 30 years after the last 
``flushing" of the reconnection ring.
It is smaller at earlier times (see Figure 4 of Shu et al.\ 2001).
For their preferred value of $S$, the maximum growth rate (at 30 yr) is seen from Figure 4 of
Shu et al.\ (2001) to reach $d a / dt \sim +0.03 \, {\rm cm} \, {\rm yr}^{-1}$ at late times
(10 years or later).
From 1-2 years, the growth rates are much smaller, $< 3 \times 10^{-4} \, {\rm cm} \, {\rm yr}^{-1}$. 

These growth rates are to be compared to the rate at which hydrogen ions in the plasma thermally 
sputter the proto-CAIs, an effect neglected by Shu et al.\ (2001). 
The density of hydrogen ions arises straightforwardly from the density of hydrogen gas Shu et al.\ (2001)
assume is trapped on field lines crossing the reconnection ring, 
$\sim 5 \times 10^{-16} \, {\rm g} \, {\rm cm}^{-3}$.
Jones (2004) gives a simple formula for the sputtering rate in a hot plasma:
$d a / dt \sim -(n_{\rm H} / 10^{10} \, {\rm cm}^{-3} ) \, {\rm yr}^{-1}$.
For $n_{\rm H} \sim 3 \times 10^{8} \, {\rm cm}^{-3}$, this means even a large 1 cm CAI will be completely
sputtered in only 30 yr.
A more detailed discussion can be found in Draine \& Salpeter (1979), who calculate that 
in a $T \sim 10^7 \, {\rm K}$ plasma, each impacting H ion yields roughly 0.02 atoms liberated
from an impacted silicate (and 0.2 atoms per impact of He ions).
Given the flux of H atoms $n_{\rm H} v_{\rm T} / 4 \sim 3 \times 10^{15} \, {\rm cm}^{-2} \, {\rm s}^{-1}$
in the ring, it is straightforward to show that particles, again, shrink at a rate
$d a / dt \sim - 0.03 \, {\rm cm} \, {\rm yr}^{-1}$.
This is competitive with the fastest growth rates of the largest particles at about 30 years, implying
that for the sticking coefficient assumed by Shu et al.\ (2001), particles do not grow faster than
they are sputtered. 
The sputtering rate is independent of particle size, and acts even when particles are small.
Thus, about 1 year after material has been flushed out of the reconnection ring, when the largest
particles are about 70 microns in radius (according to Shu et al.\ 2001), thermal sputtering
acts about 100 times faster than growth by vapor deposition.
Particles at this stage could only grow if the effective (time-averaged) sticking coefficient were $> 0.1$, 
which is implausibly high.
Neglect of thermal sputtering by Shu et al.\ (2001) is a serious oversight; inclusion of this
effect shows that particles will not survive, nor grow, in the reconnection ring. 

\subsection{Retrieval in a Magnetocentrifugal Outflow?}

Above we have argued that large particles cannot grow in the reconnection ring, because
they are likely to be sputtered before they grow, or are likely to collide fast enough to
shatter each other.
Assuming that particles do grow, and do have low relative velocities, then in principle
they could be launched in magnetocentrifugal outflows when the protostellar magnetic cycle
ebbs and the disk encroaches on the reconnection ring; but in practice it is not clear 
that particles can be launched.
Gas orbiting the protostar is launched in a magnetocentrifugal outflow when it is tied to
magnetic field lines inclined from vertical by a critical amount ($60^{\circ}$).
Ionized gas is tied to magnetic field lines (because of flux freezing) like beads on a
wire; when these wires are inclined to the vertical and spun around an axis, the beads
tied to the wire are flung outward.
Because of symmetry, magnetic field lines are exactly vertical when they penetrate the midplane
of a protoplanetary disk; gas at the midplane will not be flung outward.
Wardle \& K\"{o}nigl (1993) have examined the vertical structure of accretion disks from
which gas is being magnetocentrifugally launched.
They find that such outflows are launched only from heights $z$ above the midplane in
excess of 2 gas pressure heights $H$.

In order to be launched in a magnetocentrifugal outflow, large particles must be located
at least $2 H$ above the midplane; if they are not, they will be tied to gas that is
not moving upward and is not being flung out along field lines.
For parameters typical of the inner edge of the disk ($T = 1500 \, {\rm K}$
sound speed $2.3 \, {\rm km} \, {\rm s}^{-1}$, $\Omega = 1 \times 10^{-5} \, {\rm s}^{-1}$),
the pressure scale height is $H \sim C / \Omega = 2 \times 10^{10} \, {\rm cm}$, and particles
must reach heights $z > 4 \times 10^{10} \, {\rm cm}$ above the midplane to be launched.
The actual vertical distribution of particles as the disk encroaches on them is much smaller,
though, on the order of $w_{z} / \Omega$.
As $w_{z} < 0.03 \, {\rm km} \, {\rm s}^{-1}$ by necessity (or else proto-CAIs would shatter
on impact and never grow), their vertical distribution is limited to
$z < 3 \times 10^{8} \, {\rm cm}$, at least initially.
Without some intervening mechanism to vertically spread them, these CAIs will never be launched.
Shu et al.\ (2001) do not identify such a mechanism.

The most plausible mechanism for lofting large particles above the midplane is turbulence,
perhaps driven by a magnetorotational instability (MRI) acting at the X point, as Shu
et al.\ (2001) suggest acts to transfer gas across the X point. 
Several opposing constraints must be satisfied for this to occur.
According to the X wind model,
the magnetic diffusivity of the gas must be sufficiently high that mass can diffuse off of
field lines threading the disk, and onto field lines tied to the star; but the diffusivity
can not be so high that it suppresses the MRI generating the turbulence.
It is not clear these conditions can both be met: a magnetic diffusivity
$> 0.3 \, H^2 \, \Omega \sim 2 \times 10^{15} \, {\rm cm}^{2} \, {\rm s}^{-1}$ at the
X point will suppress the MRI (e.g., Desch 2004); on the other hand, for matter to radially diffuse a
distance $\sim 0.1 R_{x}$ in 1 yr requires a comparable diffusivity
$> 2 \times 10^{14} \, {\rm cm}^2 \, {\rm s}^{-1}$.
At any rate, detailed modeling of the X point is required before the MRI can be
invoked as a source of turbulence, let alone yield the exact turbulence needed to loft
CAI-sized particles.
In the absence of such a mechanism, the proto-CAIs in the reconnection ring will retain
whatever vertical distribution they exhibited there, and they will not be launched.

A lack of detailed modeling also hinders judgment of the last element of launching
in the magnetocentrifugal outflow, the final trajectories taken by launched CAIs.
Examples of calculated trajectories are presented in Shu et al.\ (1996) but the 
calculations on which they are based have not appeared in the refereed literature.
One conclusion about these trajectories that is probably robust is that the trajectories 
taken by specific particles are {\it highly} sensitive to their aerodynamic properties.
Shu et al.\ (1996) define a parameter $\alpha$, inversely proportional to the product a
particle's density and radius.
Particles with identical $\alpha$ will follow identical trajectories, but particles with
slightly differing $\alpha$ will follow greatly varying trajectories.
A factor of 2 variation in particle size is the difference between falling back onto
the disk at 0.2 AU, or leaving the solar system altogether.
Given this sensitivity, it is not clear that many particles would be of the right size to 
be launched on trajectories that deposit them in the 2-3 AU region. 

\section{X-Wind Model Predictions and Meteoritic Constraints}

The X-wind model, as reviewed above, has many internal inconsistencies.
It also makes predictions about the formation of chondrules and CAIs that are 
inconsistent with their petrology and other meteoritic constraints. 
Formation of SLRs in their meteoritic abundances also faces difficulties
in the context of the X-wind model.
These inconsistencies are discussed in this section.

\subsection{Chondrule Formation}

The X-wind model is inconsistent with the thermal histories of chondrule formation, constraints
on which were discussed in \S 2. 
The typical disk temperatures just outside the X point, where chondrules form in the X-wind model,
are typically $> 1160 \, {\rm K}$, far higher than the temperatures ($\approx 650 \, {\rm K}$)
require to condense primary sulfur. 
The cooling rates of chondrules in the X-wind model are $\sim 10 \, {\rm K} \, {\rm hr}^{-1}$
for all particles.  
These cooling rates match those required to produce  porphyritic chondrule textures as they
pass through their crystallization temperatures; but they are not consistent with the cooling 
rates of barred olivine chondrules, $250 - 3000 \, {\rm K} \, {\rm hr}^{-1}$. 
They also are not consistent with the much more rapid cooling rates above the liquidus, needed to 
retain volatiles such as S and Na. 
Finally, the correlation between chondrule cooling rate and the compound chondrule frequency, which
is a robust prediction of the nebular shock model (Desch \& Connolly 2002; Ciesla \& Hood 2002), is 
unexplained by the X-wind model. 

Some aspects of the chondrule formation environment in the X-wind model are consistent with constraints,
others not.  
The chondrule formation environment is not explicitly modeled within the X-wind model, but we 
can estimate the gas density.
Adopting a minimum-mass solar nebula profile (Weidenschilling 1977a), we infer a gas density 
$\approx 2 \times 10^{-6}$ at 0.05 AU, or higher if the disk mass exceeds the minimum-mass solar 
nebula mass. 
Assuming a typical solids/gas density ratio $5 \times 10^{-3}$ and a typical chondrule mass 
$\approx 3 \times 10^{-4} \, {\rm g}$, we infer a number density of chondrules 
$\approx 30 \, {\rm m}^{-3}$.
This is slightly higher but not inconsistent with the density of chondrules based on compound 
chondrule frequency and volatile retention. 
One prediction by the X-wind model about the chondrule formation environment is robust, though:
chondrules were heated near 0.1 AU and launched to the 2-3 AU region, where they joined cold
dust that had never been heated.
This is inconsistent with the presence of matrix dust that was indeed heated to high temperatures,
even condensed, in the chondrite-forming region (Scott \& Krot 2005). 
Micron-sized matrix grains launched by the X-wind are predicted to not fall back on the disk, so it
is difficult to explain the presence of such grains.
Moreover, matrix grains and chondrules within a given chondrite are chemically complementary (at 
least in their refractory lithophiles), meaning that chondrules and matrix grains are derived from
the same batch of solar-composition material. 

Finally, the X-wind, model predicts that chondrules and CAIs are formed contemporaneously, and offers
no explanation for the observed time difference $\sim 2$ Myr between CAI and chondrule formation.

\subsection{CAI Formation}

One of the successes of the X-wind model was its prediction that comets would contain
CAIs (Shu et al.\ 1996), like the inclusion {\it Inti} retrieved by the {\it STARDUST} mission 
from comet Wild 2 (Zolensky et al.\ 2006), although other physical models also predict outward
transport of CAIs in the disk (Desch 2007; Ciesla 2007).
The X-wind model is inconsistent with many other aspects of CAI formation. 
It is a robust prediction of the X-wind model that CAIs should evaporate and recondense in a very 
oxidizing environment.
According to Shu et al.\ (2001), the density of hydrogen gas in the reconnection ring
is ${\cal C}^{-1} \, \times (2 \times 10^{-16}) \,{\rm g} \, {\rm cm}^{-3}$, where 
${\cal C}$ is a dimensionless quantity near unity (see discussion before their equation 12).
Alternatively, they estimate the electron density in this region to be 
$n_{\rm e} \approx 3 \times 10^{8} \, {\rm cm}^{-3}$.
For an ionized hydrogen gas, this yields a density 
$5 \times 10^{-16} \, {\rm g} \, {\rm cm}^{-3}$, which is the value we adopt. 
In the X-wind model, proto-CAIs grow by condensation following large flares that evaporate
much of the solid material.
Following an event that evaporates all of the ferromagnesian mantle material from proto-CAIs, 
Shu et al.\ (2001) estimate (their \S 5.1) a surface density $\sim 1.6 \, {\rm g} \, {\rm cm}^{-2}$
of rocky material (presumably FeO, MgO and ${\rm SiO}_{2}$) in the gas phase.
Initially this material is confined to the volume occupied by the thin disk of proto-CAIs, but it
will thermally expand.
If it is allowed to expand along field lines more than $\sim 10^{12} \, {\rm cm}$ above the
reconnection ring, the gas will be lost to the protostar; Shu et al.\ (2001) assert that 
the gas will cool before that time.
At any rate, the very lowest density the rock vapor can have corresponds to the maximum
vertical distribution of about $10^{12} \, {\rm cm}$, which yields a density of rock vapor
$\sim (1.6 \, {\rm g} \, {\rm cm}^{-2}) / (2 \times 10^{12} \, {\rm cm})$
$\sim 1 \times 10^{-12} \, {\rm g} \, {\rm cm}^{-3}$.
That is, the mass density of heavy elements is 2000 times the density of hydrogen.
This is to be compared to the ratio in a solar-composition gas, $\sim 0.015$.
Expressed as an oxygen fugacity, it is seen that CAIs materials condense out of a gas that
is over 5 orders of magnitude more oxidizing than a solar-composition gas, i.e., with 
$f{\rm O}_{2} \approx {\rm IW} - 1$.
The high oxygen fugacity of the gas in the reconnection ring during the times when 
gas is condensing onto proto-CAIs is completely inconsistent with the barometers of
oxygen fugacity such as Ti valence states in fassaite and rh\"{o}nite, which imply 
a near solar-composition gas (Krot et al.\ 2000).
It is also inconsistent with the condensation of osbornite in some CAIs, especially in
the object known as {\it Inti} in the {\it STARDUST} sample return (Meibom et al.\ 2007);
the osbornite also must have condensed in a solar composition gas (Ebel \& Grossman 2000).
Indeed, the presence of N in the reconnection ring in the first place may itself be
problematic, as it should be quickly swept up in the funnel flow. 

\subsection{Radionuclide Production} 

The X-wind model was developed to explain the abundances of the SLRs ${}^{41}{\rm Ca}$,
${}^{26}{\rm Al}$, ${}^{53}{\rm Mn}$ and ${}^{10}{\rm Be}$ together, but in fact the model
has difficulty matching the meteoritic abundances of these SLRs.
In the X-wind model, production of $^{26}$Al without overproducing $^{41}$Ca requires that 
ferromagnesian silicate mantles surround CAI-like refractory cores, and that the two components form 
immiscible melts during heating. 
This absolute need arises because in their model ${}^{41}{\rm Ca}$ is produced by spallation of
${}^{40}{\rm Ca}$, whereas ${}^{26}{\rm Al}$ is produced from spallation of Mg. 
Without sequestration of Ca in a core, beneath a mantle $> 1 \, {\rm mm}$ thick to shield
solar energetic particles, ${}^{41}{\rm Ca}$ is consistently overproduced in the X-wind model,
relative to ${}^{26}{\rm Al}$. 
Shu et al.\ (2001) argue that Ca and Al should be sequestered in a core using theoretical 
arguments, but experiments consistently show that Ca,Al-rich silicates have a lower melting 
point than ferromagnesian silicates and do not form immiscible melts as Shu et al.\ (2001) 
describe, instead being well mixed (Simon et al.\ 2002). 

Significantly, both radionuclides are underproduced relative to ${}^{10}{\rm Be}$ in the X-wind
model.
This is because the dominant target nucleus, ${}^{16}{\rm O}$, is distributed throughout the
CAI, and because the reaction proceeds most rapidly due to higher energy
($\sim 50 \, {\rm MeV} \, {\rm nucleon}^{-1}$) solar energetic particles that can penetrate the CAI.
Gounelle et al.\ (2001) were able to marginally co-produce ${}^{26}{\rm Al}$ and ${}^{10}{\rm Be}$
using a theoretically derived rate for the reaction ${}^{3}{\rm He}({}^{24}{\rm Mg},p){}^{26}{\rm Al}$.
In fact, this reaction rate has been experimentally measured and found to be 3 times smaller than 
Gounelle et al.\ (2001) had assumed (Fitoussi et al.\ 2004), meaning that ${}^{10}{\rm Be}$ is 
overproduced by at least a factor of 3 relative to ${}^{26}{\rm Al}$ in CAIs in the X-wind model.
Recent modeling of radionuclide production in the X-wind environment confirms the overabundance
of ${}^{10}{\rm Be}$ relative to ${}^{26}{\rm Al}$ (Sahijpal \& Gupta 2009).
The discrepancy is worsened if, in fact, the majority of ${}^{10}{\rm Be}$ comes from trapped
GCRs, as advocated by Desch et al.\ (2004). 

The X-wind model is not capable of explaining the presence of ${}^{60}{\rm Fe}$ in the early solar system.
The neutron-rich isotope ${}^{60}{\rm Fe}$ is underproduced relative to other radionuclides
(e.g., ${}^{26}{\rm Al}$) by orders of magnitude (Leya et al.\ 2003; Gounelle 2006). 
In order to explain the abundance of ${}^{60}{\rm Fe}$ in the solar nebula, a separate, nucleosynthetic
source is required, probably a single nearby supernova (or a small number of nearby supernovae), 
which could have injected many other SLRs at the same time. 

The presence of ${}^{36}{\rm Cl}$ also does not appear to be explained by the X-wind model. 
Its presence in the solar nebula has been interpreted as evidence for a late stage of irradiation 
within the solar nebula, producing ${}^{36}{\rm Cl}$ by direct bombardment of target nuclei by 
energetic ions (Lin et al.\ 2005; Hsu et al.\ 2006; Jacobsen et al.\ 2009).
The X-wind model provides a natural environment for irradiation to take place, but production 
of ${}^{36}{\rm Cl}$ requires irradiation of the target nuclei S, Cl, Ar, and K.  
The 50\% condensation temperatures of all of these elements exceed 1000 K (Lodders 2003), so at 
the X-point none of these elements will condense.
If any of these elements are carried into the reconnection ring, they will quickly evaporate 
and join the funnel flow and be accreted onto the star.
Significantly, if ${}^{36}{\rm Cl}$ were created in the reconnection ring, it would fail to 
recondense following the evaporation of CAI material. 
The presence of live ${}^{36}{\rm Cl}$ in meteoritic inclusions perhaps implies irradiation, but 
only in a relatively cold environment ($< 1000 \, {\rm K}$), far cooler than the X-wind model
predicts. 
The fact that the ${}^{36}{\rm Cl}$ occurs in late-stage alteration products like sodalite
also argues against production at the same time CAI were forming.

Within the context of the X-wind model, the SLRs ${}^{10}{\rm Be}$, ${}^{41}{\rm Ca}$, ${}^{26}{\rm Al}$
and ${}^{53}{\rm Mn}$ are corproduced in their observed proportions only after making assumptions
about the behavior of CAI melts and the cross section of the ${}^{16}{\rm O}(p,x){}^{10}{\rm Be}$
reaction that are not justified.
In particular, ${}^{10}{\rm Be}$ is likely to be overproduced significantly relative to other SLRs 
in the X-wind environment. 
The X-wind model also provides no explanation for ${}^{60}{\rm Fe}$ and ${}^{36}{\rm Cl}$ in the
early solar system, and these SLRs must have a separate origin, perhaps a nearby supernova or 
irradiation in colder regions of the disk. 
It is likely that these other sources would contribute to the inventories of other SLRs as well.
This is not to rule out contributions from the X-wind, but to point out that the X-wind model must
be seen as one model among many alternatives.
We now consider the ability of alternative models to explain chondrule and CAI formation, and the
origins of the SLRs. 

\section{Alternatives to the X Wind}

The X-wind model attempted to connect three distinct problems in meteoritics
to a single astrophysical model, to advance the field toward ``an astrophysical
theory of chondrites".
The problems of chondrule formation, CAI formation, and the origins of the SLRs
are not wholly unconnected.
On the other hand, extensive petrological and cosmochemical measurements had
already led, and have continued to lead, the meteoritics community to develop 
detailed theories for each of these problems.  
We summarize these here, to provide the astrophysics community with a current
review of these fields, and to provide a comparison for the X wind model, so
that its successes and failures can be put into a proper perspective. 

\subsection{Chondrule Formation}

At this time, the leading model for chondrule formation is passage through 
nebular shock waves, in the protoplanetary disk.
The model was first proposed by Wood (1963) and subsequently developed by 
Hood \& Horanyi (1991, 1993), Connolly \& Love (1998), Hood (1998), Iida et al.\ (2001),
Desch \& Connolly (2002), Ciesla \& Hood (2002), Miura \& Nakamoto (2006), 
and Morris \& Desch (2010).  
Reviews of chondrule formation and the shock model can be found in 
Jones et al.\ (2000), Connolly \& Desch (2004), Desch et al.\ (2005), 
Hewins et al.\ (2005), and Connolly et al.\ (2006).
Two leading candidates for the source of the shocks are gravitational instabilities
that drive spiral shocks through the disk, or bow shocks around planetesimals on
eccentric orbits.
Gravitational instabilities would naturally produce large shocks at high speeds
compatible with the shock models, if the disk can be shown to be unstable 
(Boss \& Durisen 2005; Boley \& Durisen 2008).
Because instability requires a cold, massive disk, it may be delayed until
mass piles up in the disk and the disk cools; a delay of 2 Myr is not 
unreasonable.
Planetesimal bow shocks should be ubiquitous if planetesimals form early
(by some process that does not rely on chondrule formation) and Jupiter can 
pump up the eccentricities of these bodies (Hood et al.\ 2009).  
Formation of a massive Jupiter might take 2 Myr, so a delay between CAI and
chondrule formation is again not unreasonable.
The two shock models and their relative merits are discussed further by
Desch et al.\ (2005). 
In either model of chondrule formation by shocks, chondrule precursors are 
melted in the disk, at about 2-3 AU, in the presence of dust, thereby complying 
with the constraints of chondrule-matrix complementarity and the presence of 
condensate grains discussed above.
Turbulence in the disk is capable of generating regions of varying chondrule
density (Cuzzi et al.\ 2001, 2008; Teitler et al.\ 2009),
exceeding $10^2$ on lengthscales $\sim 10^4 \, {\rm km}$ (Hogan \& Cuzzi 2007;
Cuzzi et al.\ 2008).
The shock wave is presumed to advance through the disk, and individual chondrules 
would be melted in microenvironments varying in chondrule density and oxidation
state. 

The models of Desch \& Connolly (2002) and Ciesla \& Hood (2002), as well
as Morris \& Desch (2010), are in general agreement and calculate similar
thermal histories for chondrules. 
A typical case is depicted in Figure 1, for a pre-shock gas density 
$10^{-9} \, {\rm g} \, {\rm cm}^{-3}$, chondrule-to-gas mass ratio of $3.75\%$
and shock speed $8 \, {\rm km} \, {\rm s}^{-1}$.
The disk gas is presumed to be cold enough to condense S, because at the time
of chondrule formation, 2 Myr after CAI formation, the disk is in the passively 
heated protoplanetary disk stage (Chiang \& Goldreich 1997).
As the shock advances, radiation from already heated chondrules escapes to the
pre-shock region, pre-heating chondrules (perhaps forming melt that draws 
fluffy aggregates into compact spheres before the shock hits).
Peak temperatures are reached immediately after the shock hits and are
$\approx 2000 \, {\rm K}$ for these parameters.
Peak temperatures are attributable to the combination of absorption of other
chondrules' radiation, thermal exchange with the compressed, heated gas, and the 
drag heating as the chondrules equilibrated to the gas velocity.
This drag heating disappears in one aerodynamic stopping time, about 1 minute,
implying initial cooling rates $\sim 10^4 \, {\rm K} \, {\rm hr}^{-1}$.
Chondrules then cool from about 1700 K at the rates at which they pass many optical
depths from the shock front, $\sim 10 - 10^2 \, {\rm K} \, {\rm hr}^{-1}$ depending
on the density of chondrules which provide the opacity (dust is predicted to 
evaporate in the shock: Morris \& Desch 2010). 
The shock model predicts the cooling rate through the crystallization temperatures
is proportional to the chondrule density. 
The two stages of cooling and the cooling rate proportional to chondrule density are 
robust predictions unique to the shock model.

These are to be compared to the thermal histories of chondrules in the X wind model,
superimposed on Figure 1.  
Parameters for the ``revealed stage", in which $\dot{M} = 1 \times 10^{-7} \, M_{\odot} \, {\rm yr}^{-1}$,
were adopted. 
Temperatures in the X wind model are too high initially to condense S (at least in a 
near-solar composition gas), do not heat by more than a few hundred K, do not reach 
temperatures several hundred K above the liquidus, and do not exhibit two stages of 
cooling with fast initial cooling rate and slower cooling rate at lower temperatures.
The chondrules' cooling rates also are not proportional to the chondrule density. 

%

To summarize, the shock model conforms to many constraints that the X wind model does not.
It predicts thermal histories with cold initial temperature, rapid rise to the correct
peak temperatures, rapid cooling at first, then slow cooling through the crystallization 
range.
The X wind model predicts high initial temperatures, a limited temperature increase to the 
peak temperature, and a single cooling rate from the peak temperature.
The shock model predicts that chondrule cooling rates, which determine textures, are 
proportional to the chondrule density, explaining why barred olivine textures, which demand 
fast cooling rates and therefore chondrule densities, are more prevalent in compound chondrules.
The X wind model predicts no correlation of cooling rate with chondrule density, and no
correlation of chondrule texture with compound chondrule frequency. 
The shock model is consistent with formation in the disk and therefore both the presence
of condensate grains and chondrule-matrix chemical complementarity.
The X wind model would predict no correlation between chondrules and the matrix in which
they are sited, and explicitly predicts that matrix grains have never been heated.
Either of the proposed mechanisms for shocks, gravitational instability and planetesimal
bow shocks, is compatible with a 2 Myr delay between CAI and chondrule formation.
The X wind model predicts contemporaneous production of CAIs and chondrules.
The shock model makes detailed predictions about the chondrule formation environment 
and the thermal histories of chondrules; the X wind model is less detailed, but where it 
makes predictions these often fail to conform to constraints. 
The data overwhelmingly support an origin for chondrules in the disk, melted by 
nebular shocks, rather than formation in the X-wind environment. 

\subsection{CAI formation} 

The formation of CAIs is a major unsolved problem in meteoritics.
Fluffy Type A CAIs and the precursors of other, melted, CAIs contain refractory
minerals that condense at high temperatures (Grossman 2002; MacPherson 2003).
Barometers of oxygen fugacity constrain this gas to be as reducing as one of
solar composition.  
These factors point to condensation in the solar nebula, at a stage when it was 
very hot, implying formation at an early time and/or location closer to the Sun. 
At temperatures $\approx \, 1500 - 1650 \, {\rm K}$, for example, hibonites
and other Ca,Al-rich minerals in CAIs would condense, but ferromagnesian silicates 
would not (Lodders 2003).
Models of the structure of protoplanetary disks that include realistic opacity, 
convection and viscous heating predict temperatures $> 1400 \, {\rm K}$ only 
inside about 0.5 AU, even if the mass accretion rate through the disk is as high 
as $\dot{M} = 10^{-7} \, M_{\odot} \, {\rm yr}^{-1}$, a stage that can only last
for $\sim 0.5 \, {\rm Myr}$ or less. 
Formation of CAIs during this restricted time of the disk's evolution is consistent
with the inferred spread in CAI ages $\approx 0.4 \, {\rm Myr}$, derived from Al-Mg
systematics (MacPherson et al.\ 1995; Kita et al.\ 2005, 2010; Shahar \& Young 2007).

The main objection to this straightforward interpretation is the so-called ``CAI storage
problem", the perceived inability of solids to remain in the protoplanetary disk for 
the $\sim 2 \, {\rm Myr}$ needed so that CAIs can join chondrules in chondrites. 
Aerodynamic drag, in particular, is expected to cause CAIs to spiral in toward the
Sun on timescales $\sim 10^5 \, {\rm yr}$ (Weidenschilling 1977b). 
Cuzzi et al.\ (2003) have shown, however, that while the majority of CAIs may migrate
inward on $10^5 - 10^6 \, {\rm yr}$ timescales, turbulence causes CAIs to diffuse outward
as well on the same timescales. 
This model predicts that smaller CAIs should diffuse outward more effectively than larger
particles, explaining the greater prevalence of Type A CAIs relative to the larger Type B 
CAIs.
Within the context of the same model, Cuzzi et al.\ (2005a, 2005b) have also shown 
that CAIs experience high temperatures for the long ($\sim 10^4 - 10^5 \, {\rm yr}$
timescales needed for elements to diffuse across the so-called Wark-Lovering rims
observed around many CAIs. 
The igneous textures of most CAIs are potentially explained by passage through nebular 
shocks, in much the same manner as chondrules are presumably melted. 
The peak temperatures and cooling rates are consistent with this scenario. 
It is not clear whether shocks that melted CAIs would have been identical to the 
ones that melted chondrules, or perhaps were just due to the same mechanism but 
acting in a different environment.
It is also not clear that such shocks could have acted at the times needed to melt CAIs.
In principle, however, shocks acting in the disk over many Myr could explain the
igneous textures of most CAIs. 
Thus, storage of CAIs in the disk is not only allowed by disk models, but may be 
necessary to explain their mineralogy and textures.

The scenario outlined above is consistent with the mineralogy of CAIs, especially
formation of CAIs in a reducing gas.
In contrast, 
The X-wind model predicts that CAIs should condense in their own rock vapor, 
devoid of almost all ${\rm H}_{2}$ gas, and is not consistent at all with the low 
oxygen fugacity recorded by CAIs during their formation.
The scenario outlined above also is consistent with an early formation of CAIs over a 
short interval, whereas the X wind model predicts that CAIs should form continuously 
over many Myr. 

\subsection{Short-lived Radionuclides}

\subsubsection{Iron 60 and Others}

Essentially the only explanation for the presence of ${}^{60}{\rm Fe}$ in the early solar 
system is that it was injected into the solar nebula by one nearby supernova, or a small
number of nearby supernovae (Goswami et al.\ 2005; Meyer \& Zinner 2006; Wadhwa et al.\ 2007).
Irradiation within the solar nebula or in the X-wind environment fails to produce the
observed initial abundance of this neutron-rich isotope, by many orders of magnitude
(Leya et al.\ 2003; Gounelle 2006).
An external stellar nucleosynthetic source is demanded.  
An AGB star has been suggested as the source (Wasserburg et al.\ 1994, 
1995, 1996, 1998), but isotopic evidence argues against an AGB star origin (Wadhwa et al.\ 2007), 
as well as the fact that a nearby AGB star at the time and place of the solar system's formation
is exceedingly improbable (Kastner \& Myers 1994; Ouellette et al.\ 2009). 
The only plausible stellar source is a core-collapse supernova, because massive stars
($> 20 \, M_{\odot}$) can evolve off the main sequence and explode as supernovae in 
$< 10 \, {\rm Myr}$, before they disperse from their birth clusters. 
It is currently debated whether the solar nebula's ${}^{60}{\rm Fe}$ originated in a single 
supernova, less than 1 pc away, or in many supernovae several parsecs distant.
Constraining which scenario applies is important for determine what radionuclides are 
injected along with ${}^{60}{\rm Fe}$. 
For a single supernova, the distances must be nearby, less than several parsecs (Looney et al.\
2006).
Injection by a single supernova, into the Sun's protoplanetary disk (Chevalier 2000), has been 
advocated by Ouellette et al.\ (2005, 2007, 2010), who show that sufficient ${}^{60}{\rm Fe}$ 
could be injected into an extant protoplanetary disk if it were a few $\times 0.1 \, {\rm pc}$ 
from an isotropically exploding supernova, or up to a few parsecs away from a supernova as clumpy 
as the ejecta in the Cassiopeia A supernova remnant (see also Looney et al.\ 2006).
Gounelle \& Meibom (2008) and Gaidos et al.\ (2009) have argued that young ($< 1 \, {\rm Myr}$
old) disks $< 1 \, {\rm pc}$ from a supernova are rare, occurring with $< 1\%$ probability;
Ouellette et al.\ (2010) likewise calculate a low probability $\sim 1\%$ for a disk at 2 pc
to be struck by ejecta. 
Additionally, injection into the disk requires much of the ejecta to condense into dust grains
before encountering the disk: simulations show $< 1\%$ of the intercepted gas ejecta is injected 
into a disk (Ouellette et al.\ 2007).
Injection of gas into a molecular cloud, instead of a disk, in principle can occur as far as a 
few parsecs (Looney et al.\ 2006; Gaidos et al.\ 2009), but here again the injection efficiency 
of gas ejecta is $\sim 1\%$ (Boss et al.\ 2010).
Recent models of supernova shock-triggered collapse by Boss \& Keiser (2010) do exhibit
shock fronts that are thinner and denser than those previously considered, and may allow
for greater injection efficiencies.  
At this point, injection from a single supernova into either a protoplanetary disk or 
molecular cloud are viable models, although they might entail improbable circumstances.

Gounelle et al.\ (2009) have proposed that the gas from which the Sun formed was
contaminated by several dozen core-collapse supernovae, then swept up into a molecular 
cloud several Myr before the solar system formed.
Their ``Supernova Propagation and Cloud Enrichment" (SPACE) model invokes an astrophysical 
setting like the Scorpius-Centaurus star-forming region, in which massive stars have triggered
collapse of nearby molecular clouds (either by winds or supernova shocks), triggering a
new round of massive star formation and supernovae, (cf. Preibisch \& Zinnecker 1999).
In their model, Gounelle et al.\ (2009) computed an average value 
${}^{60}{\rm Fe} / {}^{56}{\rm Fe} \approx 3 \times 10^{-6}$ in a molecular cloud 
over a 10-20 Myr span, assuming a half-life of 1.5 Myr; updating the half-life to 
2.3 Myr (Rugel et al.\ 2009) potentially could raise the ${}^{60}{\rm Fe}$ abundance 
by an order of magnitude, assuming the molecular cloud takes 10 Myr to form the Sun.
A weakness of the model is that the supernova ejecta is assumed to mix into the 
swept-up material with 100\% efficiency.
Simulations of supernova ejecta interacting with protoplanetary disks (Ouellette et al.\
2007) and molecular clouds (Boss et al.\ 2010) typically find mixing efficiencies
$\sim 1\%$.
Gounelle et al.\ (2009) argue for high mixing efficiencies on the basis of simulations
of the thermal instability in interstellar shocks that do suggest high mixing ratios
(Koyama \& Inutsuka 2002; Audit \& Henebelle 2010).
These latter simulations, it should be noted, involve shock speeds of only a few 
$\times 10 \, {\rm km} \, {\rm s}^{-1}$, for which the post-shock temperature is
$< 10^{4} \, {\rm K}$ and is consistent with a thermally unstable gas. 
The shock speeds associated with supernova ejecta less than a few parsecs from 
the explosion center are necessarily $\sim 10^{3} \, {\rm km} \, {\rm s}^{-1}$,
and in these shocks the post-shock gas is too hot to cool effectively. 
We expect the mixing efficiency of supernova ejecta with swept-up gas to be 
closer to 1\% than 100\%, and consider the mixing efficiency to be an unresolved issue
with the SPACE model. 

Assuming the validity of either model, we can estimate the abundances of other
radionuclides injected along with ${}^{60}{\rm Fe}$, especially the shortest lived
of the SLRs, ${}^{41}{\rm Ca}$, ${}^{36}{\rm Cl}$, ${}^{26}{\rm Al}$, ${}^{10}{\rm Be}$ 
and ${}^{53}{\rm Mn}$. 
Neither model is capable of explaining ${}^{10}{\rm Be}$, which is not created
by stellar nucleosynthesis; the case of ${}^{10}{\rm Be}$ is considered separately
below. 
As for the others, it has been demonstrated that a single supernova can inject 
the other radionuclides in the observed meteoritic proportions, provided the 
progenitor is $> 20 \, M_{\odot}$ so that when it undergoes core collapse it may
result in the ``faint supernova" type in which the innermost layers fall back onto
the core (Umeda \& Nomoto 2002, 2005; Nomoto et al.\ 2006; Tominaga et al.\ 2007).
Because essentially all of the ${}^{53}{\rm Mn}$ in a supernova is produced in the
innermost $3 \, M_{\odot}$ (Nomoto et al.\ 2006), fallback of ejecta reduces the 
${}^{53}{\rm Mn} / {}^{26}{\rm Al}$ ratio in the ejecta by orders of magnitude,
resulting in the observed meteoritic proportions of ${}^{41}{\rm Ca}$, ${}^{26}{\rm Al}$, 
${}^{60}{\rm Fe}$ and ${}^{53}{\rm Mn}$ in the ejecta, assuming a reasonable 1 Myr 
delay before isotopic closure (Takigawa et al.\ 2008).
The abundance of ${}^{36}{\rm Cl}$ in the early solar system appears to be too high
to be explained by injection from a single supernova (see discussion in Hsu et al.\
2006). 
Injection of material from a single nearby supernova, either into the disk or into
the Sun's molecular cloud core, can simultaneously explain the abundances of the
other shortest-lived radionuclides ${}^{41}{\rm Ca}$, ${}^{26}{\rm Al}$, 
${}^{60}{\rm Fe}$ and ${}^{53}{\rm Mn}$, if the progenitor was a massive star 
experiencing fallback.  

Within the context of the SPACE model, injection of ${}^{60}{\rm Fe}$ from 
multiple supernovae may yield the meteoritic ${}^{26}{\rm Al} / {}^{60}{\rm Fe}$
ratio in the solar nebula, but cannot explain the abundances of ${}^{41}{\rm Ca}$ 
and ${}^{53}{\rm Mn}$.
The SPACE model does not lead to significant quantities of SLRs with half-lives
$< 1 \, {\rm Myr}$, because of the long timescales (10-20 Myr) associated with 
the formation of the molecular cloud, so ${}^{41}{\rm Ca}$ and ${}^{36}{\rm Cl}$ would
be significantly underproduced.
This underproduction is inconsistent with studies that indicate a correlation between
${}^{26}{\rm Al}$ and ${}^{41}{\rm Ca}$ (Sahijpal \& Goswami 1998), unless ${}^{26}{\rm Al}$ 
is not derived primarily from these multiple supernovae. 
Likewise, the SPACE model unavoidably and significantly {\it over}produces ${}^{53}{\rm Mn}$
(Gounelle et al.\ 2009). 
Models of supernova ejecta generally show a ${}^{53}{\rm Mn} / {}^{60}{\rm Fe}$ ratio
10 - 100 times larger than the solar nebula ratio inferred from meteorites (Goswami \&
Vanhala 2000; Wadhwa et al.\ 2007; Sahijpal \& Soni 2006).
This general trend does not apply to ejecta from a single supernova, if the supernova's
progenitor was $> 20 \, M_{\odot}$ and experienced fallback, but considering the average ejecta 
of dozens of supernovae of various masses, this outcome appears inevitable. 

To summarize, ${}^{60}{\rm Fe}$ cannot be formed by the X-wind model and requires an
external supernova source. 
The multiple supernovae in the SPACE model of Gounelle et al.\ (2009) explains the 
the abundance of ${}^{60}{\rm Fe}$ in the early solar system, assuming that mixing 
efficiencies approach unity.
Production of ${}^{41}{\rm Ca}$, ${}^{36}{\rm Cl}$ and ${}^{26}{\rm Al}$ in the X-wind
environment would not conflict with production of ${}^{60}{\rm Fe}$ in the SPACE model,
but the SPACE model inevitably and significantly overproduces ${}^{53}{\rm Mn}$,
making it incompatible with the X-wind model for SLR production, which also contributes
to ${}^{53}{\rm Mn}$. 
Because of this severe overproduction of ${}^{53}{\rm Mn}$ relative to ${}^{60}{\rm Fe}$,
and because we expect mixing ratios of supernova ejecta should be $\sim 1\%$, we disfavor 
the SPACE model as the source of the solar system's ${}^{60}{\rm Fe}$ and other SLRs. 
This suggests strongly that the source of the solar nebula's ${}^{60}{\rm Fe}$ was instead 
a single core-collapse supernova with progenitor mass $> 20 \, M_{\odot}$, that experienced 
fallback onto the core.
Such a supernova would have underproduced ${}^{36}{\rm Cl}$, but could simultaneously explain 
the observed abundances of ${}^{41}{\rm Ca}$, ${}^{26}{\rm Al}$, ${}^{60}{\rm Fe}$ and 
${}^{53}{\rm Mn}$ (Takigawa et al.\ 2008), without contributions from multiple supernovae 
of a previous generation of star formation.
Because a single supernova is favored source for ${}^{60}{\rm Fe}$, and because this
scenario can explain all of the SLRs (except ${}^{10}{\rm Be}$) that the X-wind model 
produces, significant contributions of these SLRs from the X-wind most likely can be
excluded. 

\subsubsection{The Special Case of Beryllium 10}
  
Since evidence for ${}^{10}{\rm Be}$ in the solar nebula was discovered (McKeegan et al.\ 2000),
it has been used to support the X-wind model.
Because this SLR is not produced in supernovae, Gounelle et al.\ (2001) called it a potential 
``smoking gun" for the X-wind model.
However, the data point to an origin for ${}^{10}{\rm Be}$ that is distinct from ${}^{26}{\rm Al}$
and the other SLRs.
Marhas et al.\ (2002) analyzed a variety of meteoritic components thought to form early
in the solar nebula, including a so-called FUN (fractionation and unknown nuclear effects) CAI, 
as well as hibonites. 
They found evidence for ${}^{10}{\rm Be}$ in samples with firm upper limits on initial ${}^{26}{\rm Al}$,
and concluded that ${}^{10}{\rm Be}$ was not correlated with ${}^{26}{\rm Al}$, and the two
SLRs were ``decoupled," having separate origins, a conclusion supported by subsequent studies
(Ushikubo et al.\ 2006; Srinivasan et al.\ 2007). 
In addition, the initial abundances of ${}^{10}{\rm Be}$ in a variety of samples are 
remarkably uniform. 
Desch et al.\ (2004) reviewed the dozen or so measurements up to that date and found them to
all cluster in a range ${}^{10}{\rm Be} / {}^{9}{\rm Be}$ $\approx \, 0.45 - 1.8 \times 10^{-3}$.
More measurements have been made since then, all of which again cluster in the same range
(Marhas et al.\ 2002; MacPherson et al.\ 2003; Ushikubo et al.\ 2006; Chaussidon et al.\ 2006;
Srinivasan et al.\ 2007; Liu et al.\ 2007).
These data strongly suggest that the source of ${}^{10}{\rm Be}$ not only was distinct from
the source of ${}^{26}{\rm Al}$ and other SLRs, but pre-dated the solar system. 

Desch et al.\ (2004) interpret these data to mean that most of the ${}^{10}{\rm Be}$ was
inherited from the interstellar medium, as ${}^{10}{\rm Be}$ GCRs
that were slowed and trapped in the Sun's molecular cloud core as it collapsed.
They calculated the rate at which such low-energy ($< 10 \, {\rm MeV} \, {\rm nucleon}^{-1}$)
GCRs were trapped in the Sun's cloud core, accounting for magnetic focusing and mirroring,
and computed an initial ratio in the solar system of $^{10}{\rm Be} / {}^{9}{\rm Be} = 1.1 \times 10^{-3}$.
Other SLRs are not predicted to derive from this mechanism (Desch et al.\ 2004). 
To the extent that any fraction of the ${}^{10}{\rm Be}$ in the solar nebula comes from
a source other than the X-wind, it exacerbates the problems of overproduction of 
${}^{10}{\rm Be}$ in the X-wind, relative to other SLRs (\S 5.3). 
If Desch et al.\ (2004) are correct in their interpretation that nearly all the ${}^{10}{\rm Be}$
came from trapped GCRs, it effectively rules out the X-wind model for SLR production. 
Because of its important consequences for the X-wind model, the model of Desch et al.\ (2004)
has been questioned; here we address these criticisms.

Desch et al.\ (2004) predicted that the ${}^{10}{\rm Be} / {}^{9}{\rm Be}$ ratio was initially
homogeneous within the solar nebula, as it represents material that was trapped in the molecular
cloud core.  
In truth, fewer GCRs would reach and be stopped in the center of cloud core, so the 
${}^{10}{\rm Be} / {}^{9}{\rm Be}$ ratio would not have been completely homogeneous at
this stage; it is difficult to judge the degree of heterogeneity at this stage, although
it is probably less than a factor of 2. 
At any rate, it is presumed that such heterogeneities are erased as the cloud core continues
to collapse into a protostar and disk, and the prediction of homogeneity of ${}^{10}{\rm Be}$
probably is robust. 
Gounelle (2006) claimed that the variations in inferred initial 
${}^{10}{\rm Be} / {}^{9}{\rm Be}$ ratios point to a non-homogeneous distribution of ${}^{10}{\rm Be}$.
Likewise, Liu et al.\ (2007) analyzed platy hibonites from CM chondrites and found one with 
an initial ratio ${}^{10}{\rm Be} / {}^{9}{\rm Be} = 5.5 \pm 1.4 \times 10^{-4}$ which, they claimed,
was statistically significantly lower than the average values.
Because platy hibonites are believed to be older than other components, this lower value is
not attributed to decay of ${}^{10}{\rm Be}$ over time, implying that ${}^{10}{\rm Be}$ was
spatially heterogeneous. 
Notably, though, Ushikubo et al.\ (2006) also measured platy hibonites from the CM2 chondrite 
Murchison and the CO3 chondrite Kainsaz, and inferred higher initial values in similar samples,
${}^{10}{\rm Be} / {}^{9}{\rm Be} = 1.8 \pm 0.4 \times 10^{-3}$.
We choose to interpret the range of inferred initial ${}^{10}{\rm Be} / {}^{9}{\rm Be}$ ratios 
as clustering about a uniform value, within the experimental uncertainties. 
Clearly, further analyses will determine whether observed variations reflect true nebular
heterogeneities or differences in experimental techniques.

Gounelle (2006) also criticized many assumptions and other aspects of the Desch et al.\ (2004)
model. 
First, they disputed the long cloud core collapse time $\sim 10 \, {\rm Myr}$ used in the
main simulation of Desch et al.\ (2004), implying that since observed collapse times of
molecular cloud cores are $\approx 0.3 - 1.6 \, {\rm Myr}$, (Lee \& Myers 1999), that
perhaps Desch et al.\ (2004) overestimated the ${}^{10}{\rm Be}$ by a factor $\approx 10$.
In fact, it is clear from Figure 3 of Desch et al.\ (2004) that the ${}^{10}{\rm Be} / {}^{9}{\rm Be}$
quickly saturates to values $\sim 1 \times 10^{-3}$, so longer collapse times do {\it not} 
lead to higher ${}^{10}{\rm Be} / {}^{9}{\rm Be}$ ratios.
In fact, Desch et al.\ (2004) explored the sensitivity of ${}^{10}{\rm Be}$ abundance to
magnetic field strength and therefore collapse time (their Figure 4).
They found ${}^{10}{\rm Be} / {}^{9}{\rm Be} \approx 1 \times 10^{-3}$ even for parameters
that lead to collapse times $< 1 \, {\rm Myr}$.
Gounelle (2006) also criticized the assumption of Desch et al.\ (2004) that the GCR flux was
a factor of 2 higher 4.6 Gyr ago than today, calling it ``ad hoc". 
In fact, as explained by Desch et al.\ (2004), the GCR flux scales with the supernova rate,
which scales with the star formation rate, which is well known to be decreasing over Galactic
history. 
The GCR flux was definitely higher in the past than today, by a factor roughly $1.5 - 2.5$ higher
than today (Desch et al.\ 2004). 
Gounelle (2006) also criticized the fact that the simulations of Desch \& Mouschovias (2001)
used by Desch et al.\ (2004) formed a $1 \, M_{\odot}$ star from a $45 \, M_{\odot}$ cloud.
We point out that the observed star formation efficiency is similarly low (Ward-Thompson et al.\
2007), and that the exact cloud structure is somewhat irrelevant: ${}^{10}{\rm Be}$ GCRs 
will be trapped in collapsing cores, as demonstrated by Desch et al.\ (2004), as they transition
from low densities transparent to low-energy GCRs to high densities opaque to GCRs, passing
through surface densities $\sim 10^{-2} \, {\rm g} \, {\rm cm}^{-2}$.
This is true regardless of the details of the larger structures, because they are largely
transparent to such GCRs. 
Other objections raised by Gounelle (2006), e.g., relating to the importance of magnetic mirroring, 
are addressed directly by Desch et al.\ (2004)
The objections of Gounelle (2006) are readily refuted, and we consider the model of
Desch et al.\ (2004) to be valid. 

Beryllium 10 is known to be decoupled from the other SLRs and to have a separate source.
The uniformity of the inferred initial ${}^{10}{\rm Be} / {}^{9}{\rm Be}$ ratios around 
a value $\approx 1 \times 10^{-3}$ strongly suggests an origin before the formation of 
the protoplanetary disk. 
The model of Desch et al.\ (2004) predicts ${}^{10}{\rm Be} / {}^{9}{\rm Be} \approx 1 \times 10^{-3}$
due to trapping of low-energy ${}^{10}{\rm Be}$ GCRs as the Sun's molecular cloud core
contracts and becomes opaque to such GCRs. 
To the extent that ${}^{10}{\rm Be}$ in the early solar system can be attributed to trapped 
GCRs, then the contributions to ${}^{10}{\rm Be}$ must be significantly reduced or even 
excluded, effectively ruling out significant contributions to the SLRs from the X-wind. 

\section{Conclusions}

The X-wind model was originally developed to explain the dynamics of bipolar outflows from
protostars (Shu et al.\ 1994a,b, 1995; Najita \& Shu 1994; Ostriker \& Shu 1995). 
It remains a viable model for protostellar jets, although not the only one: ``disk wind" models,
in which the magnetocentrifugal outflows are launched from 0.1 - 1 AU, rather than from $< 0.1$ AU,
also exist (Wardle \& K\"{o}nigl 1993; K\"{o}nigl \& Pudritz 2000; Pudritz et al.\ 2007).
Observational evidence from the rotation of protostellar jets tends to favor disk wind models
(Bacciotti et al.\ 2002; Anderson 2003; Coffey et al.\ 2004, 2007; Woitas et al.\ 2005), 
and at this time the evidence for X-wind models in particular
is not conclusive.

In a series of papers (Shu et al.\ 1996, 1997, 2001; Gounelle et al.\ 2001), the X-wind model was 
applied to three fundamental problems in meteoritics: the formation of chondrules, the formation of CAIs,
and the origin of the SLRs.
Progress toward an {\it astrophysical} theory of chondrites was sought. 
In this paper, we have shown that the X-wind model is not applicable to the formation of chondrules,
to the formation of CAIs, nor the origin of the SLRs.
We have demonstrated that the model itself has internal inconsistencies.  
It also makes predictions about chondrule and CAI formation at odds with experimental constraints.
In regard to the SLRs, it does not satisfactorily explain the coproduction of ${}^{10}{\rm Be}$,
${}^{26}{\rm Al}$, ${}^{41}{\rm Ca}$ and ${}^{53}{\rm Mn}$, and it leaves unexplained the 
source of ${}^{60}{\rm Fe}$ and ${}^{36}{\rm Cl}$.

The internal inconsistencies can be summarized as follows.
First, material is brought to the reconnection ring only because of accretion, yet the heating
caused by this accretion was neglected in the X-wind model.
When it is included, the model predicts temperatures too high for most silicate material to exist.
This is consistent with astronomical observations, which also show no evidence for solids at the
X point.
Second, the X-wind model assumes rather arbitrarily that a fraction $F \sim 0.01$ of all solid 
material falls out of the funnel flow and into the reconnection ring.  
This factor is not determined from first principles, and our own calculations presented here 
show that essentially all solids brought in from the disk will remain entrained in the funnel 
flow and accreted on the star.
Third, the X-wind model asserts that particles falling from the funnel flow will join a 
geometrically thin ``reconnection ring".
In fact, particles leaving the funnel flow are likely to enter the reconnection ring with 
velocities comparable to the Keplerian orbital velocity there, $> 100 \, {\rm km} \, {\rm s}^{-1}$,
with significant orbital inclinations.
It is not clear how these inclinations would be damped so particles could join the reconnection 
ring. 
Also, particles already in the reconnection ring would experience shattering collisions with
incoming particles, at a rate sufficient to prevent particle growth in the reconnection ring.
Fourth, the X-wind model neglects thermal sputtering by the plasma in the reconnection ring.
We have shown that thermal sputtering will prevent growth of particles in the reconnection ring.
Fifth, the X-wind model necessarily posits that CAIs and chondrules, formed in the reconnection
ring, lie very close to the disk midplane ($< 3 \times 10^{9} \, {\rm cm}$), yet particles 
must be far from the midplane ($> 4 \times 10^{10} \, {\rm cm}$) to be launched in a 
magnetocentrifugal outflow.
Vertical diffusion of particles is not modeled.
The MRI is invoked, but the magnetic diffusivity needed to allow gas to diffuse across the X point 
is close to the limit at which the MRI is suppressed.
Sixth, the trajectories of particles launched in the magnetocentrifugal outflow are not explicitly
modeled.
It does seem clear, though, that the mechanism is extremely sensitive to the size of the particles,
implying that only a small fraction of the material could be launched. 

Ignoring these internal inconsistencies, the X-wind model makes a number of predictions about 
chondrules that are inconsistent with constraints on their origins. 
The thermal histories of chondrules are experimentally constrained by measurements of elemental 
and isotopic fractionation, and by chemical zoning and textures.
The X-wind model does not allow chondrules to form from material containing primary S, as the 
starting temperatures are too high.  
It does not explain the very high peak temperatures of chondrules, nor the rapid cooling from the
peak.
It also predicts that all CAIs and chondrules melted in the X-wind will cool at $10 \, {\rm K} \, {\rm hr}^{-1}$,
which is not consistent with the cooling rates of barred olivine and some other chondrules,
$\sim 10^{3} \, {\rm K} \, {\rm hr}^{-1}$.
The observed correlation between compound chondrule frequency and textural type is also not
predicted by the X-wind model.  
Very importantly, the X-wind model predicts that within a chondrite the chondrules are
formed at $< 0.1 \, {\rm AU}$ and the matrix grains at $\approx 2 - 3 \, {\rm AU}$, and that there
should be no correlation between their compositions.
This directly contradicts the observed chondrule-matrix chemical complementarity. 
Finally, the X-wind model predicts contemporaneous formation of chondrules and CAIs, which 
is contradicted by Pb-Pb dating and Al-Mg systematics, which show a 2 Myr age difference. 

The X-wind model also makes a number of predictions about CAI formation. 
The assumption of a refractory Ca,Al-rich core surrounded by a ferromagnesian silicate mantle 
(necessary to prevent substantial overproduction of ${}^{41}{\rm Ca}$) is not supported by
observed behaviors of CAI melts. 
Also, because CAIs are explicitly assumed to grow due to vapor recondensation, the oxygen fugacity 
of the X-wind environment will be that of rock vapor itself; hydrogen and other volatile phases would 
be accreted by the funnel flow onto the star.
This oxygen fugacity is orders of magnitude too oxidizing to be consistent with oxygen barometers
of CAI formation, which routinely indicate a gas of solar composition.
The discovery of osbornite in the CAI-like {\it Stardust} sample {\it Inti}, and in CAIs of
Isheyevo, likewise strongly indicate a gas of solar composition for the formation environments
of these particular inclusions, and not an X-wind environment.

The X-wind model also makes a number of predictions about the production of SLRs.
In the context of the X-wind model, even for the most favorable parameters (Gounelle et al.\ 2001),
${}^{10}{\rm Be}$ is overproduced, given that the cross section 
${}^{24}{\rm Mg}({}^{3}{\rm He},p){}^{26}{\rm Al}$ 
is measured to be 3 times smaller than Gounelle et al.\ (2001) assumed (Fitoussi et al.\ 2004).
The overproduction of ${}^{10}{\rm Be}$ is more profound to the extent that ${}^{10}{\rm Be}$
has an external origin, such as trapped GCRs (Desch et al.\ 2004). 
In the context of the X-wind model, the only way to avoid severe overproduction of ${}^{41}{\rm Ca}$
is if almost all the Ca in the CAI were sequestered in a core, surrounded by a silicate mantle 
$\sim 1 \, {\rm cm}$ thick.  
As Simon et al.\ (2002) point out, real CAI melts do not form immiscible liquids that would 
segregate in this way. 
Despite the likelihood that ${}^{36}{\rm Cl}$ in the solar nebula was created by irradiation, 
the X-wind environment is too hot for either the target nuclei or ${}^{36}{\rm Cl}$ to condense. 
Finally, the X-wind model cannot explain the existence of ${}^{60}{\rm Fe}$ in the solar nebula,
because this neutron-rich isotope is not sufficiently produced by spallation.

The problems of the X-wind model are even starker in the face of the viable alternatives that 
exist in the literature. 
Chondrule formation is explained in great detail by melting in nebular shocks. 
This model is consistent with the detailed thermal histories of chondrules, their observed 
correlation with compound chondrule frequency, and chondrule-matrix complementarity.
Formation of CAIs in the disk, during an earlier stage of disk evolution where the mass
accretion rates were higher, is consistent with an earlier formation of CAIs than chondrules,
with the solar oxygen fugacity of their formation environment, and allows some CAIs to remain
unmelted.
Finally, because ${}^{60}{\rm Fe}$ is not produced significantly in the X-wind environment, 
its source must be one or more nearby core-collapse supernovae.
The overproduction of ${}^{53}{\rm Mn}$ relative to ${}^{60}{\rm Fe}$ appears to exclude 
multiple supernovae. 
Injection of material from a single, nearby core-collapse supernova is broadly consistent
and can explain simultaneously the abundances of ${}^{41}{\rm Ca}$, ${}^{26}{\rm Al}$,
${}^{60}{\rm Fe}$ and ${}^{53}{\rm Mn}$ (Takigawa et al.\ 2008).
Neither one nor several supernova, nor the X-wind model, appear capable of explaining 
the high inferred initial abundance of ${}^{36}{\rm Cl}$, which may demand a separate
origin in a late stage of irradiation in the early solar system.
Supernova nucleosynthesis does not produce ${}^{10}{\rm Be}$, but this SLR is known to be 
decoupled from ${}^{26}{\rm Al}$ and the other SLRs.
A unique origin as trapped GCRs qualitatively and quantitatively explains
its near-uniform abundance ${}^{10}{\rm Be} / {}^{9}{\rm Be} \sim 10^{-3}$ in a variety of 
meteoritic inclusions (Desch et al.\ 2004).
Objections by Gounelle (2006) to the model of Desch et al.\ (2004) are readily refuted.
The origins of the SLRs are still unknown and are the focus of ongoing research; but the
working hypothesis of trapped ${}^{10}{\rm Be}$ GCRs and injection from a single supernova
with fallback appears more viable than the X-wind model plus multiple supernovae for ${}^{60}{\rm Fe}$. 
In short, viable and more plausible alternative models exist for all the meteoritic components
the X-wind model purports to explain. 

The X-wind model makes assumptions that are internally inconsistent.
The X-wind model makes predictions about the formation of chondrules and CAIs and the 
production of SLRs that are contradicted by experimental constraints. 
Better alternative models exist to explain the formation of chondrules and CAIs and
the production of SLRs.
We conclude the X-wind model is irrelevant to the problems of chondrule formation, CAI
formation, or the creation of short-lived radionuclides.

\acknowledgements
S.~J.~D.\ gratefully acknowledges the support for this work made available by NASA's Origins 
of Solar Systems Program, grant NNG06GI65G, and by the NASA Astrobiology Institute.


\newpage

\begin{figure}[ht]
\plotone{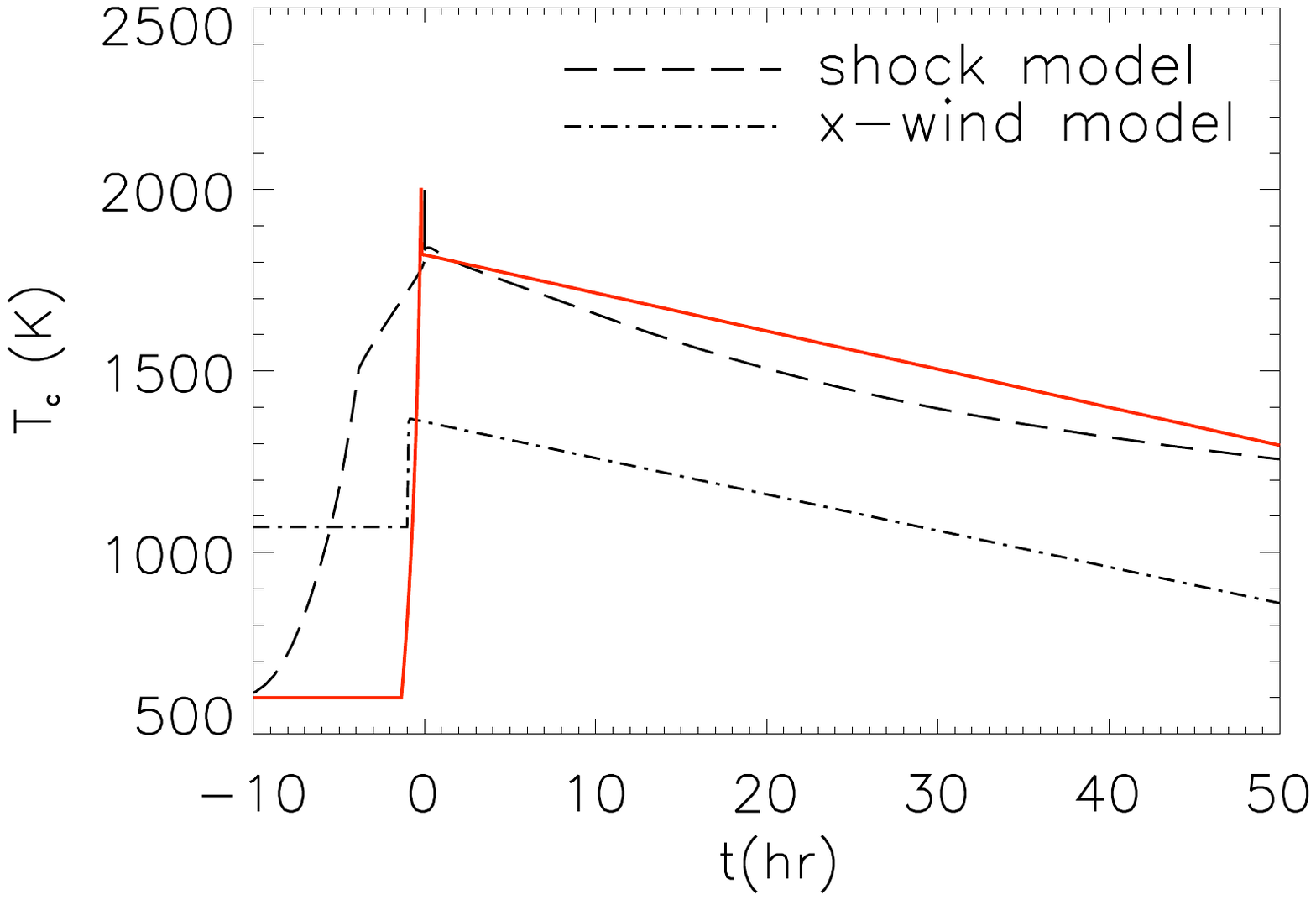}
\vspace{-2in} 
\caption{Chondrule thermal histories as inferred from experimental constraints 
           (solid curve), as predicted by the shock model [from Morris \& Desch (2010)] 
           (dashed curve), and as predicted by the X-wind model during the ``revealed stage" 
           [adapted from Shu et al.\ (1996, 2001)] (dashed-dot curve).  Chondrules in the 
           X-wind model start too hot to condense S from a solar-composition gas, fail to 
           reach the necessary peak temperatures, and show no rapid cooling from the peak
           that is needed to retain Na.  Except for the prediction of an extended period of 
           pre-shock heating, the shock model conforms well to all the constraints. }
\end{figure}

\end{document}